\documentclass[sigconf]{acmart}
\usepackage{siunitx}
\usepackage{colortbl}
\usepackage{tikz}
\usepackage{pgfplots}

\usepackage{acronym}

\providecommand{\dotdiv}{
\mathbin{
    \vphantom{+}
    \text{
      \mathsurround=0pt 
      \ooalign{
        \noalign{\kern-.35ex}
        \hidewidth$\smash{\cdot}$\hidewidth\cr 
        \noalign{\kern.35ex}
        $-$\cr 
      }%
    }%
  }%
}
\usepackage{amsmath}
\usepackage{centernot}
\usepackage{algpseudocode}
\usepackage{algorithm}
\usepackage{multirow}
\usepackage{subcaption}

\usepackage{natbib}

\theoremstyle{definition}

\definecolor{mm}{RGB}{127,201,127}
\definecolor{cl}{RGB}{204,0,0}
\definecolor{jgs}{RGB}{253,192,134}
\definecolor{td}{RGB}{56,108,176}
\definecolor{lcs}{RGB}{153,50,204}
\definecolor{darkgreen}{RGB}{0,102,51}

\definecolor{notcorrect}{RGB}{220,20,60}

\definecolor{gray}{RGB}{220,220,220}

\newcommand{\topicset}[0]{\ensuremath{T}}

\newcommand{\groundtruthmap}[0]{\ensuremath{\textrm{GT}}}

\newcommand{\allpreceq}[0]{\ensuremath{\sqsubseteq}}

\AtBeginDocument{%
  \providecommand\BibTeX{{%
    \normalfont B\kern-0.5em{\scshape i\kern-0.25em b}\kern-0.8em\TeX}}}

\copyrightyear{2021} 
\acmYear{2021} 
\setcopyright{none} 
\acmConference[CIKM '21]{Proceedings of the 30th ACM International Conference on Information and Knowledge Management}{November 1--5, 2021}{Virtual Event, QLD, Australia}
\acmBooktitle{Proceedings of the 30th ACM International Conference on Information and Knowledge Management (CIKM '21), November 1--5, 2021, Virtual Event, QLD, Australia}
\acmDOI{10.1145/3459637.3482287}
\acmISBN{978-1-4503-8446-9/21/11}

\begin{document}

\fancyhead{}
\title{Principled Multi-Aspect Evaluation Measures of Rankings}

\author{Maria Maistro}
\email{mm@di.ku.dk}
\affiliation{%
\institution{University of Copenhagen}
\country{Denmark}
}

\author{Lucas Chaves Lima}
\email{lcl@di.ku.dk}
\affiliation{%
\institution{University of Copenhagen}
\country{Denmark}
}

\author{Jakob Grue Simonsen}
\email{simonsen@di.ku.dk}
\affiliation{%
\institution{University of Copenhagen}
\country{Denmark}
}

\author{Christina Lioma}
\email{c.lioma@di.ku.dk}
\affiliation{%
\institution{University of Copenhagen}
\country{Denmark}
}

\renewcommand{\shortauthors}{Maistro, M., et al.}

\begin{abstract}
Information Retrieval evaluation has traditionally focused on defining principled ways of assessing the relevance of a ranked list of documents with respect to a query. Several methods extend this type of evaluation beyond relevance, making it possible to evaluate different aspects of a document ranking (e.g., relevance, usefulness, or credibility) using a single measure (\textit{multi-aspect evaluation}). However, these methods either are (i) tailor-made for specific aspects and do not extend to other types or numbers of aspects, or (ii) have theoretical anomalies, e.g. assign maximum score to a ranking where all documents are labelled with the lowest grade with respect to all aspects (e.g., not relevant, not credible, etc.). 

We present a theoretically principled multi-aspect evaluation method that can be used for any number, and any type, of aspects. 
A thorough empirical evaluation using up to $5$ aspects and a total of $425$ runs officially submitted to $10$ TREC tracks shows that our method is more discriminative than the state-of-the-art and overcomes theoretical limitations of the state-of-the-art. 
\end{abstract}

\ccsdesc[500]{Information systems~Information retrieval}
\ccsdesc[500]{Information systems~Evaluation of retrieval results}
\ccsdesc[500]{Information systems~Retrieval effectiveness}

\keywords{Evaluation, ranking, multiple aspects, partial order}

\maketitle
\acrodef{3G}[3G]{Third Generation Mobile System}
\acrodef{5S}[5S]{Streams, Structures, Spaces, Scenarios, Societies}
\acrodef{AAAI}[AAAI]{Association for the Advancement of Artificial Intelligence}
\acrodef{AAL}[AAL]{Annotation Abstraction Layer}
\acrodef{AAM}[AAM]{Automatic Annotation Manager}
\acrodef{ACLIA}[ACLIA]{Advanced Cross-Lingual Information Access}
\acrodef{ACM}[ACM]{Association for Computing Machinery}
\acrodef{ADSL}[ADSL]{Asymmetric Digital Subscriber Line}
\acrodef{ADUI}[ADUI]{ADministrator User Interface}
\acrodef{AIP}[AIP]{Archival Information Package}
\acrodef{AJAX}[AJAX]{Asynchronous JavaScript Technology and \acs{XML}}
\acrodef{ALU}[ALU]{Aritmetic-Logic Unit}
\acrodef{AMUSID}[AMUSID]{Adaptive MUSeological IDentity-service}
\acrodef{ANOVA}[ANOVA]{ANalysis Of VAriance}
\acrodef{ANSI}[ANSI]{American National Standards Institute}
\acrodef{AP}[AP]{Average Precision}
\acrodef{APC}[APC]{AP Correlation}
\acrodef{API}[API]{Application Program Interface}
\acrodef{AR}[AR]{Address Register}
\acrodef{AS}[AS]{Annotation Service}
\acrodef{ASAP}[ASAP]{Adaptable Software Architecture Performance}
\acrodef{ASI}[ASI]{Annotation Service Integrator}
\acrodef{ASL}[ASL]{Achieved Significance Level}
\acrodef{ASM}[ASM]{Annotation Storing Manager}
\acrodef{ASR}[ASR]{Automatic Speech Recognition}
\acrodef{ASUI}[ASUI]{ASsessor User Interface}
\acrodef{ATIM}[ATIM]{Annotation Textual Indexing Manager}
\acrodef{AUC}[AUC]{Area Under the ROC Curve}
\acrodef{AUI}[AUI]{Administrative User Interface}
\acrodef{AWARE}[AWARE]{Assessor-driven Weighted Averages for Retrieval Evaluation}
\acrodef{BANKS-I}[BANKS-I]{Browsing ANd Keyword Searching I}
\acrodef{BANKS-II}[BANKS-II]{Browsing ANd Keyword Searching II}
\acrodef{bpref}[bpref]{Binary Preference}
\acrodef{BNF}[BNF]{Backus and Naur Form}
\acrodef{BRICKS}[BRICKS]{Building Resources for Integrated Cultural Knowledge Services}
\acrodef{CAM}[CAM]{Convex Aggregating Measure}
\acrodef{CAN}[CAN]{Content Addressable Netword}
\acrodef{CAS}[CAS]{Content-And-Structure}
\acrodef{CBSD}[CBSD]{Component-Based Software Developlement}
\acrodef{CBSE}[CBSE]{Component-Based Software Engineering}
\acrodef{CB-SPE}[CB-SPE]{Component-Based \acs{SPE}}
\acrodef{CD}[CD]{Collaboration Diagram}
\acrodef{CD}[CD]{Compact Disk}
\acrodef{CENL}[CENL]{Conference of European National Librarians}
\acrodef{CIDOC CRM}[CIDOC CRM]{CIDOC Conceptual Reference Model}
\acrodef{CIR}[CIR]{Current Instruction Register}
\acrodef{CIRCO}[CIRCO]{Coordinated Information Retrieval Components Orchestration}
\acrodef{CG}[CG]{Cumulated Gain}
\acrodef{CL}[CL]{Curriculum Learning}
\acrodef{CLEF1}[CLEF]{Cross-Language Evaluation Forum}
\acrodef{CLEF}[CLEF]{Conference and Labs of the Evaluation Forum}
\acrodef{CLIR}[CLIR]{Cross Language Information Retrieval}
\acrodef{CM}[CM]{Continuation Methods}
\acrodef{CMS}[CMS]{Content Management System}
\acrodef{CMT}[CMT]{Campaign Management Tool}
\acrodef{CNR}[CNR]{Italian National Council of Research}
\acrodef{CO}[CO]{Content-Only}
\acrodef{COD}[COD]{Code On Demand}
\acrodef{CODATA}[CODATA]{Committee on Data for Science and Technology}
\acrodef{COLLATE}[COLLATE]{Collaboratory for Annotation Indexing and Retrieval of Digitized Historical Archive Material}
\acrodef{CP}[CP]{Characteristic Pattern}
\acrodef{CPE}[CPE]{Control Processor Element}
\acrodef{CPU}[CPU]{Central Processing Unit}
\acrodef{CQL}[CQL]{Contextual Query Language}
\acrodef{CRP}[CRP]{Cumulated Relative Position}
\acrodef{CRUD}[CRUD]{Create--Read--Update--Delete}
\acrodef{CS}[CS]{Characteristic Structure}
\acrodef{CSM}[CSM]{Campaign Storing Manager}
\acrodef{CSS}[CSS]{Cascading Style Sheets}
\acrodef{CU}[CU]{Control Unit}
\acrodef{CUI}[CUI]{Client User Interface}
\acrodef{CV}[CV]{Cross-Validation}
\acrodef{DAFFODIL}[DAFFODIL]{Distributed Agents for User-Friendly Access of Digital Libraries}
\acrodef{DAO}[DAO]{Data Access Object}
\acrodef{DARE}[DARE]{Drawing Adequate REpresentations}
\acrodef{DARPA}[DARPA]{Defense Advanced Research Projects Agency}
\acrodef{DAS}[DAS]{Distributed Annotation System}
\acrodef{DB}[DB]{DataBase}
\acrodef{DBMS}[DBMS]{DataBase Management System}
\acrodef{DC}[DC]{Dublin Core}
\acrodef{DCG}[DCG]{Discounted Cumulated Gain}
\acrodef{DCMI}[DCMI]{Dublin Core Metadata Initiative}
\acrodef{DCV}[DCV]{Document Cut--off Value}
\acrodef{DD}[DD]{Deployment Diagram}
\acrodef{DDC}[DDC]{Dewey Decimal Classification}
\acrodef{DDS}[DDS]{Direct Data Structure}
\acrodef{DF}[DF]{Degrees of Freedom}
\acrodef{DFI}[DFI]{Divergence From Independence}
\acrodef{DFR}[DFR]{Divergence From Randomness}
\acrodef{DHT}[DHT]{Distributed Hash Table}
\acrodef{DI}[DI]{Digital Image}
\acrodef{DIKW}[DIKW]{Data, Information, Knowledge, Wisdom}
\acrodef{DIL}[DIL]{\acs{DIRECT} Integration Layer}
\acrodef{DiLAS}[DiLAS]{Digital Library Annotation Service}
\acrodef{DIRECT}[DIRECT]{Distributed Information Retrieval Evaluation Campaign Tool}
\acrodef{DKMS}[DKMS]{Data and Knowledge Management System}
\acrodef{DL}[DL]{Digital Library}
\acrodefplural{DL}[DL]{Digital Libraries}
\acrodef{DLMS}[DLMS]{Digital Library Management System}
\acrodef{DLOG}[DL]{Description Logics}
\acrodef{DLS}[DLS]{Digital Library System}
\acrodef{DLSS}[DLSS]{Digital Library Service System}
\acrodef{DM}[DM]{Data Mining}
\acrodef{DO}[DO]{Digital Object}
\acrodef{DOI}[DOI]{Digital Object Identifier}
\acrodef{DOM}[DOM]{Document Object Model}
\acrodef{DoMDL}[DoMDL]{Document Model for Digital Libraries}
\acrodef{DP}[DP]{Discriminative Power}
\acrodef{DPBF}[DPBF]{Dynamic Programming Best-First}
\acrodef{DR}[DR]{Data Register}
\acrodef{DRIVER}[DRIVER]{Digital Repository Infrastructure Vision for European Research}
\acrodef{DTD}[DTD]{Document Type Definition}
\acrodef{DVD}[DVD]{Digital Versatile Disk}
\acrodef{EAC-CPF}[EAC-CPF]{Encoded Archival Context for Corporate Bodies, Persons, and Families}
\acrodef{EAD}[EAD]{Encoded Archival Description}
\acrodef{EAN}[EAN]{International Article Number}
\acrodef{ECD}[ECD]{Enhanced Contenty Delivery}
\acrodef{ECDL}[ECDL]{European Conference on Research and Advanced Technology for Digital Libraries}
\acrodef{EDM}[EDM]{Europeana Data Model}
\acrodef{EG}[EG]{Execution Graph}
\acrodef{ELDA}[ELDA]{Evaluation and Language resources Distribution Agency}
\acrodef{ELRA}[ELRA]{European Language Resources Association}
\acrodef{EM}[EM]{Expectation Maximization}
\acrodef{EMMA}[EMMA]{Extensible MultiModal Annotation}
\acrodef{EPROM}[EPROM]{Erasable Programmable \acs{ROM}}
\acrodef{EQNM}[EQNM]{Extended Queueing Network Model}
\acrodef{ER}[ER]{Entity--Relationship}
\acrodef{ERR}[ERR]{Expected Reciprocal Rank}
\acrodef{ETL}[ETL]{Extract-Transform-Load}
\acrodef{FAST}[FAST]{Flexible Annotation Service Tool}
\acrodef{FIFO}[FIFO]{First-In / First-Out}
\acrodef{FIRE}[FIRE]{Forum for Information Retrieval Evaluation}
\acrodef{FN}[FN]{False Negative}
\acrodef{FNR}[FNR]{False Negative Rate}
\acrodef{FOAF}[FOAF]{Friend of a Friend}
\acrodef{FORESEE}[FORESEE]{FOod REcommentation sErvER}
\acrodef{FP}[FP]{False Positive}
\acrodef{FPR}[FPR]{False Positive Rate}
\acrodef{GIF}[GIF]{Graphics Interchange Format}
\acrodef{GIR}[GIR]{Geografic Information Retrieval}
\acrodef{GAP}[GAP]{Graded Average Precision}
\acrodef{GLM}[GLM]{General Linear Model}
\acrodef{GLMM}[GLMM]{General Linear Mixed Model}
\acrodef{GMAP}[GMAP]{Geometric Mean Average Precision}
\acrodef{GoP}[GoP]{Grid of Points}
\acrodef{GPRS}[GPRS]{General Packet Radio Service}
\acrodef{gP}[gP]{Generalized Precision}
\acrodef{gR}[gR]{Generalized Recall}
\acrodef{gRBP}[gRBP]{Graded Rank-Biased Precision}
\acrodef{GTIN}[GTIN]{Global Trade Item Number}
\acrodef{GUI}[GUI]{Graphical User Interface}
\acrodef{GW}[GW]{Gateway}
\acrodef{HCI}[HCI]{Human Computer Interaction}
\acrodef{HDS}[HDS]{Hybrid Data Structure}
\acrodef{HIR}[HIR]{Hypertext Information Retrieval}
\acrodef{HIT}[HIT]{Human Intelligent Task}
\acrodef{HITS}[HITS]{Hyperlink-Induced Topic Search}
\acrodef{HTML}[HTML]{HyperText Markup Language}
\acrodef{HTTP}[HTTP]{HyperText Transfer Protocol}
\acrodef{HSD}[HSD]{Honestly Significant Difference}
\acrodef{IA-ERR}[IA-ERR]{Intent Aware Expected Reciprocal Rank}
\acrodef{ICA}[ICA]{International Council on Archives}
\acrodef{ICSU}[ICSU]{International Council for Science}
\acrodef{IDF}[IDF]{Inverse Document Frequency}
\acrodef{IDS}[IDS]{Inverse Data Structure}
\acrodef{IEEE}[IEEE]{Institute of Electrical and Electronics Engineers}
\acrodef{IEI}[IEI]{Istituto della Enciclopedia Italiana fondata da Giovanni Treccani}
\acrodef{IETF}[IETF]{Internet Engineering Task Force}
\acrodef{IMS}[IMS]{Information Management System}
\acrodef{IMSPD}[IMS]{Information Management Systems Research Group}
\acrodef{indAP}[indAP]{Induced Average Precision}
\acrodef{infAP}[infAP]{Inferred Average Precision}
\acrodef{INEX}[INEX]{INitiative for the Evaluation of \acs{XML} Retrieval}
\acrodef{INS-M}[INS-M]{Inverse Set Data Model}
\acrodef{INTR}[INTR]{Interrupt Register}
\acrodef{IP}[IP]{Internet Protocol}
\acrodef{IPSA}[IPSA]{Imaginum Patavinae Scientiae Archivum}
\acrodef{IR}[IR]{Information Retrieval}
\acrodef{IRON}[IRON]{Information Retrieval ON}
\acrodef{IRON2}[IRON$^2$]{Information Retrieval On aNNotations}
\acrodef{IRON-SAT}[IRON-SAT]{\acs{IRON} - Statistical Analysis Tool}
\acrodef{IRS}[IRS]{Information Retrieval System}
\acrodef{ISAD(G)}[ISAD(G)]{International Standard for Archival Description (General)}
\acrodef{ISBN}[ISBN]{International Standard Book Number}
\acrodef{ISIS}[ISIS]{Interactive SImilarity Search}
\acrodef{ISJ}[ISJ]{Interactive Searching and Judging}
\acrodef{ISO}[ISO]{International Organization for Standardization}
\acrodef{ITU}[ITU]{International Telecommunication Union }
\acrodef{ITU-T}[ITU-T]{Telecommunication Standardization Sector of \acs{ITU}}
\acrodef{IV}[IV]{Information Visualization}
\acrodef{JAN}[JAN]{Japanese Article Number}
\acrodef{JDBC}[JDBC]{Java DataBase Connectivity}
\acrodef{JMB}[JMB]{Java--Matlab Bridge}
\acrodef{JPEG}[JPEG]{Joint Photographic Experts Group}
\acrodef{JSON}[JSON]{JavaScript Object Notation}
\acrodef{JSP}[JSP]{Java Server Pages}
\acrodef{JTE}[JTE]{Java-Treceval Engine}
\acrodef{KDE}[KDE]{Kernel Density Estimation}
\acrodef{KLD}[KLD]{Kullback-Leibler Divergence}
\acrodef{KLAPER}[KLAPER]{Kernel LAnguage for PErformance and Reliability analysis}
\acrodef{LAM}[LAM]{Libraries, Archives, and Museums}
\acrodef{LAM2}[LAM]{Logistic Average Misclassification}
\acrodef{LAN}[LAN]{Local Area Network}
\acrodef{LD}[LD]{Linked Data}
\acrodef{LEAF}[LEAF]{Linking and Exploring Authority Files}
\acrodef{LIDO}[LIDO]{Lightweight Information Describing Objects}
\acrodef{LIFO}[LIFO]{Last-In / First-Out}
\acrodef{LM}[LM]{Language Model}
\acrodef{LMT}[LMT]{Log Management Tool}
\acrodef{LOD}[LOD]{Linked Open Data}
\acrodef{LODE}[LODE]{Linking Open Descriptions of Events}
\acrodef{LpO}[LpO]{Leave-$p$-Out}
\acrodef{LRM}[LRM]{Local Relational Model}
\acrodef{LRU}[LRU]{Last Recently Used}
\acrodef{LS}[LS]{Lexical Signature}
\acrodef{LSM}[LSM]{Log Storing Manager}
\acrodef{LtR}[LtR]{Learning to Rank}
\acrodef{LUG}[LUG]{Lexical Unit Generator}
\acrodef{MA}[MA]{Mobile Agent}
\acrodef{MA}[MA]{Moving Average}
\acrodef{MACS}[MACS]{Multilingual ACcess to Subjects}
\acrodef{MADCOW}[MADCOW]{Multimedia Annotation of Digital Content Over the Web}
\acrodef{MAD}[MAD]{Mean Assessed Documents}
\acrodef{MADP}[MADP]{Mean Assessed Documents Precision}
\acrodef{MADS}[MADS]{Metadata Authority Description Standard}
\acrodef{MAP}[MAP]{Mean Average Precision}
\acrodef{MAP-IA}[MAP-IA]{Intent Aware Mean Average Precision}
\acrodef{MARC}[MARC]{Machine Readable Cataloging}
\acrodef{MATTERS}[MATTERS]{MATlab Toolkit for Evaluation of information Retrieval Systems}
\acrodef{MDA}[MDA]{Model Driven Architecture}
\acrodef{MDD}[MDD]{Model-Driven Development}
\acrodef{MEM}[MEM]{Maximum Entropy Method}
\acrodef{METS}[METS]{Metadata Encoding and Transmission Standard}
\acrodef{MIDI}[MIDI]{Musical Instrument Digital Interface}
\acrodef{MIME}[MIME]{Multipurpose Internet Mail Extensions}
\acrodef{ML}[ML]{Machine Learning}
\acrodef{MLIA}[MLIA]{MultiLingual Information Access}
\acrodef{MM}[MM]{Multidimensional Measure}
\acrodef{MMU}[MMU]{Memory Management Unit}
\acrodef{MODS}[MODS]{Metadata Object Description Schema}
\acrodef{MOF}[MOF]{Meta-Object Facility}
\acrodef{MP}[MP]{Markov Precision}
\acrodef{MPEG}[MPEG]{Motion Picture Experts Group}
\acrodef{MRD}[MRD]{Machine Readable Dictionary}
\acrodef{MRF}[MRF]{Markov Random Field}
\acrodef{MS}[MS]{Mean Squares}
\acrodef{MSAC}[MSAC]{Multilingual Subject Access to Catalogues}
\acrodef{MSE}[MSE]{Mean Square Error}
\acrodef{MT}[MT]{Machine Translation}
\acrodef{MV}[MV]{Majority Vote}
\acrodef{MVC}[MVC]{Model-View-Controller}
\acrodef{NACSIS}[NACSIS]{NAtional Center for Science Information Systems}
\acrodef{NAP}[NAP]{Network processors Applications Profile}
\acrodef{NCP}[NCP]{Normalized Cumulative Precision}
\acrodef{nCG}[nCG]{Normalized Cumulated Gain}
\acrodef{nCRP}[nCRP]{Normalized Cumulated Relative Position}
\acrodef{nCT}[nCT]{Normalized Cube Test}
\acrodef{anDCG}[$\alpha$-NDCG]{$\alpha$-Normalized Discounted Cumulated Gain}
\acrodef{nDCG}[NDCG]{Normalized Discounted Cumulated Gain}
\acrodef{nDE}[nDE]{Normalized Discounted Error}
\acrodef{NESTOR}[NESTOR]{NEsted SeTs for Object hieRarchies}
\acrodef{NEXI}[NEXI]{Narrowed Extended XPath I}
\acrodef{NGRE}{Normalised Global Rank Error}
\acrodef{NII}[NII]{National Institute of Informatics}
\acrodef{NISO}[NISO]{National Information Standards Organization}
\acrodef{NIST}[NIST]{National Institute of Standards and Technology}
\acrodef{NLP}[NLP]{Natural Language Processing}
\acrodef{NLRE}{Normalised Local Rank Error}
\acrodef{NN}[NN]{Neural Network}
\acrodef{NP}[NP]{Network Processor}
\acrodef{NR}[NR]{Normalized Recall}
\acrodef{NS-M}[NS-M]{Nested Set Model}
\acrodef{NTCIR}[NTCIR]{NII Testbeds and Community for Information access Research}
\acrodef{nWCS}[nWCS]{Normalised Weighted Cumulative Score}
\acrodef{OAI}[OAI]{Open Archives Initiative}
\acrodef{OAI-ORE}[OAI-ORE]{Open Archives Initiative Object Reuse and Exchange}
\acrodef{OAI-PMH}[OAI-PMH]{Open Archives Initiative Protocol for Metadata Harvesting}
\acrodef{OAIS}[OAIS]{Open Archival Information System}
\acrodef{OC}[OC]{Operation Code}
\acrodef{OCLC}[OCLC]{Online Computer Library Center}
\acrodef{OMG}[OMG]{Object Management Group}
\acrodef{OO}[OO]{Object Oriented}
\acrodef{OODB}[OODB]{Object-Oriented \acs{DB}}
\acrodef{OODBMS}[OODBMS]{Object-Oriented \acs{DBMS}}
\acrodef{OPAC}[OPAC]{Online Public Access Catalog}
\acrodef{OQL}[OQL]{Object Query Language}
\acrodef{ORP}[ORP]{Open Relevance Project}
\acrodef{OSIRIS}[OSIRIS]{Open Service Infrastructure for Reliable and Integrated process Support}
\acrodef{P}[P]{Precision}
\acrodef{P2P}[P2P]{Peer-To-Peer}
\acrodef{PA}[PA]{Performance Analysis}
\acrodef{PAMT}[PAMT]{Pool-Assessment Management Tool}
\acrodef{PASM}[PASM]{Pool-Assessment Storing Manager}
\acrodef{PC}[PC]{Program Counter}
\acrodef{PCP}[PCP]{Pre-Commercial Procurement}
\acrodef{PCR}[PCR]{Peripherical Command Register}
\acrodef{PDA}[PDA]{Personal Digital Assistant}
\acrodef{PDF}[PDF]{Probability Density Function}
\acrodef{PDR}[PDR]{Peripherical Data Register}
\acrodef{PIR}[PIR]{Personalized Information Retrieval}
\acrodef{POI}[POI]{\acs{PURL}-based Object Identifier}
\acrodef{PoS}[PoS]{Part of Speech}
\acrodef{PPE}[PPE]{Programmable Processing Engine}
\acrodef{PREFORMA}[PREFORMA]{PREservation FORMAts for culture information/e-archives}
\acrodef{PRIMAmob-UML}[PRIMAmob-UML]{mobile \acs{PRIMA-UML}}
\acrodef{PRIMA-UML}[PRIMA-UML]{PeRformance IncreMental vAlidation in \acs{UML}}
\acrodef{PROM}[PROM]{Programmable \acs{ROM}}
\acrodef{PROMISE}[PROMISE]{Participative Research labOratory  for Multimedia and Multilingual Information Systems Evaluation}
\acrodef{pSQL}[pSQL]{propagate \acs{SQL}}
\acrodef{PUI}[PUI]{Participant User Interface}
\acrodef{PURL}[PURL]{Persistent \acs{URL}}
\acrodef{QA}[QA]{Question Answering}
\acrodef{QoS-UML}[QoS-UML]{\acs{UML} Profile for QoS and Fault Tolerance}
\acrodef{QPA}[QPA]{Query Performance Analyzer}
\acrodef{R}[R]{Recall}
\acrodef{RAM}[RAM]{Random Access Memory}
\acrodef{RAMM}[RAM]{Random Access Machine}
\acrodef{RBO}[RBO]{Rank-Biased Overlap}
\acrodef{RBP}[RBP]{Rank-Biased Precision}
\acrodef{RBU}[RBU]{Rank-Biased Utility}
\acrodef{RDBMS}[RDBMS]{Relational \acs{DBMS}}
\acrodef{RDF}[RDF]{Resource Description Framework}
\acrodef{REST}[REST]{REpresentational State Transfer}
\acrodef{REV}[REV]{Remote Evaluation}
\acrodef{RFC}[RFC]{Request for Comments}
\acrodef{RIA}[RIA]{Reliable Information Access}
\acrodef{RMSE}[RMSE]{Root Mean Square Error}
\acrodef{RMT}[RMT]{Run Management Tool}
\acrodef{ROM}[ROM]{Read Only Memory}
\acrodef{ROMIP}[ROMIP]{Russian Information Retrieval Evaluation Seminar}
\acrodef{RoMP}[RoMP]{Rankings of Measure Pairs}
\acrodef{RoS}[RoS]{Rankings of Submitted runs}
\acrodef{RP}[RP]{Relative Position}
\acrodef{RR}[RR]{Reciprocal Rank}
\acrodef{RSM}[RSM]{Run Storing Manager}
\acrodef{RST}[RST]{Rhetorical Structure Theory}
\acrodef{RT-UML}[RT-UML]{\acs{UML} Profile for Schedulability, Performance and Time}
\acrodef{SA}[SA]{Software Architecture}
\acrodef{SAL}[SAL]{Storing Abstraction Layer}
\acrodef{SAMT}[SAMT]{Statistical Analysis Management Tool}
\acrodef{SAN}[SAN]{Sistema Archivistico Nazionale}
\acrodef{SASM}[SASM]{Statistical Analysis Storing Manager}
\acrodef{SD}[SD]{Sequence Diagram}
\acrodef{SE}[SE]{Search Engine}
\acrodef{SEBD}[SEBD]{Convegno Nazionale su Sistemi Evoluti per Basi di Dati}
\acrodef{SFT}[SFT]{Satisfaction--Frustration--Total}
\acrodef{SIL}[SIL]{Service Integration Layer}
\acrodef{SIP}[SIP]{Submission Information Package}
\acrodef{SKOS}[SKOS]{Simple Knowledge Organization System}
\acrodef{SM}[SM]{Software Model}
\acrodef{SME}[SME]{Statistics--Metrics-Experiments}
\acrodef{SMART}[SMART]{System for the Mechanical Analysis and Retrieval of Text}
\acrodef{SoA}[SoA]{Service-oriented Architectures}
\acrodef{SOA}[SOA]{Strength of Association}
\acrodef{SOAP}[SOAP]{Simple Object Access Protocol}
\acrodef{SOM}[SOM]{Self-Organizing Map}
\acrodef{SPARQL}[SPARQL]{Simple Protocol and RDF Query Language}
\acrodef{SPE}[SPE]{Software Performance Engineering}
\acrodef{SPINA}[SPINA]{Superimposed Peer Infrastructure for iNformation Access}
\acrodef{SPLIT}[SPLIT]{Stemming Program for Language Independent Tasks}
\acrodef{SPOOL}[SPOOL]{Simultaneous Peripheral Operations On Line}
\acrodef{SQL}[SQL]{Structured Query Language}
\acrodef{SR}[SR]{Sliding Ratio}
\acrodef{SR}[SR]{Status Register}
\acrodef{SRU}[SRU]{Search/Retrieve via \acs{URL}}
\acrodef{SS}[SS]{Sum of Squares}
\acrodef{SSTF}[SSTF]{Shortest Seek Time First}
\acrodef{STAR}[STAR]{Steiner-Tree Approximation in Relationship graphs}
\acrodef{STON}[STON]{STemming ON}
\acrodef{SVM}[SVM]{Support Vector Machine}
\acrodef{TAC}[TAC]{Text Analysis Conference}
\acrodef{TBG}[TBG]{Time-Biased Gain}
\acrodef{TCP}[TCP]{Transmission Control Protocol}
\acrodef{TEL}[TEL]{The European Library}
\acrodef{TERRIER}[TERRIER]{TERabyte RetrIEveR}
\acrodef{TF}[TF]{Term Frequency}
\acrodef{TFR}[TFR]{True False Rate}
\acrodef{TLD}[TLD]{Top Level Domain}
\acrodef{TME}[TME]{Topics--Metrics-Experiments}
\acrodef{TN}[TN]{True Negative}
\acrodef{TO}[TO]{Transfer Object}
\acrodef{TOMA}[TOMA]{Total Order Multi-Aspect}
\acrodef{TP}[TP]{True Positve}
\acrodef{TPR}[TPR]{True Positive Rate}
\acrodef{TRAT}[TRAT]{Text Relevance Assessing Task}
\acrodef{TREC}[TREC]{Text REtrieval Conference}
\acrodef{TRECVID}[TRECVID]{TREC Video Retrieval Evaluation}
\acrodef{TTL}[TTL]{Time-To-Live}
\acrodef{UCD}[UCD]{Use Case Diagram}
\acrodef{UDC}[UDC]{Universal Decimal Classification}
\acrodef{uGAP}[uGAP]{User-oriented Graded Average Precision}
\acrodef{UI}[UI]{User Interface}
\acrodef{UML}[UML]{Unified Modeling Language}
\acrodef{UMT}[UMT]{User Management Tool}
\acrodef{UMTS}[UMTS]{Universal Mobile Telecommunication System}
\acrodef{UoM}[UoM]{Utility-oriented Measurement}
\acrodef{uRBP}[uRBP]{Understandability-biased Rank-Biased Precision}
\acrodef{UPC}[UPC]{Universal Product Code}
\acrodef{URI}[URI]{Uniform Resource Identifier}
\acrodef{URL}[URL]{Uniform Resource Locator}
\acrodef{URN}[URN]{Uniform Resource Name}
\acrodef{USM}[USM]{User Storing Manager}
\acrodef{VA}[VA]{Visual Analytics}
\acrodef{VAIRE}[VAIR\"{E}]{Visual Analytics for Information Retrieval Evaluation}
\acrodef{VATE}[VATE$^2$]{Visual Analytics Tool for Experimental Evaluation}
\acrodef{VIRTUE}[VIRTUE]{Visual Information Retrieval Tool for Upfront Evaluation}
\acrodef{VD}[VD]{Virtual Document}
\acrodef{VDM}[VDM]{Visual Data Mining}
\acrodef{VIAF}[VIAF]{Virtual International Authority File}
\acrodef{VL}[VL]{Visual Language}
\acrodef{VoIP}[VoIP]{Voice over IP}
\acrodef{VS}[VS]{Visual Sentence}
\acrodef{W3C}[W3C]{World Wide Web Consortium}
\acrodef{WAN}[WAN]{Wide Area Network}
\acrodef{WHAM}[WHAM]{Weighted Harmonic Mean Aggregating Measure}
\acrodef{WHO}[WHO]{World Health Organization}
\acrodef{WLAN}[WLAN]{Wireless \acs{LAN}}
\acrodef{WP}[WP]{Work Package}
\acrodef{WS}[WS]{Web Services}
\acrodef{WSD}[WSD]{Word Sense Disambiguation}
\acrodef{WSDL}[WSDL]{Web Services Description Language}
\acrodef{WWW}[WWW]{World Wide Web}
\acrodef{XMI}[XMI]{\acs{XML} Metadata Interchange}
\acrodef{XML}[XML]{eXtensible Markup Language}
\acrodef{XPath}[XPath]{XML Path Language}
\acrodef{XSL}[XSL]{eXtensible Stylesheet Language}
\acrodef{XSL-FO}[XSL-FO]{\acs{XSL} Formatting Objects}
\acrodef{XSLT}[XSLT]{\acs{XSL} Transformations}
\acrodef{YAGO}[YAGO]{Yet Another Great Ontology}
\acrodef{YASS}[YASS]{Yet Another Suffix Stripper}
\section{Introduction}
\label{sec:introduction}

\textit{Multi-aspect evaluation} is a task in \ac{IR} evaluation where the ranked list of documents returned by an \ac{IR} system in response to a query is assessed in terms of not only relevance, but also other \textit{aspects} (or dimensions) 
such as credibility or usefulness. 
Generally, there are two ways to conduct multi-aspect evaluation: (1) evaluate each aspect separately using any appropriate single-aspect evaluation measure (e.g., \acs{AP}, \acs{nDCG}, F1), and then aggregate the scores across all aspects into a single score; or (2) evaluate all aspects at the same time using any appropriate multi-aspect evaluation measure \cite{AmigoEtAl2018,LiomaEtAl2017,TangEtAl2017}. An advantage of the aggregating option (1) 
is that it is easy to implement using evaluation measures that are readily available and well-understood in the community. 
Its disadvantage is that it is not guaranteed that all aspects will have similar distributions of labels, and aggregating across wildly different distributions can give odd results~\cite{LiomaEtAl2019}. 
The second way of doing multi-aspect evaluation is to use a single multi-aspect evaluation measure. The problem here is that few such evaluation measures exist, and most of them are defined for specific aspects and do not generalise to other types/numbers of aspects (see $\S$\ref{sec:related}).

Motivated by the above, we \textbf{contribute} a novel multi-aspect evaluation method that works with any type and number of aspects, 
and 
avoids the above problems. Given a ranked list, 
where documents are labelled with multiple aspects, our method, \acf{TOMA} evaluation, first defines a preferential order (formally \textit{weak order relation}) among documents with multiple aspect labels, and then aggregates the document labels across aspects to obtain a ranking of aggregated aspect labels, which 
can be evaluated by any single-aspect evaluation measure, such as \ac{nDCG} or \ac{AP}. Simply put, instead of evaluating each aspect separately and then aggregating their scores, we first aggregate the aspect labels and then evaluate the ranked list of documents. We do this in a way that provides several degrees of freedom: our method can be used with any number and type of aspects, can be instantiated with any binary or graded, set-based or rank-based evaluation measure, and can accommodate any granularity in the importance of each aspect or label, but still ensures, by definition, that the preference order among multi-aspect documents is not violated, and that the final measure score will meet some common requirements, i.e., the minimum (worst) score being $0$ and the maximum (perfect) score being $1$. We validate this empirically ($\S$\ref{sec:ex_setup}) and theoretically ($\S$\ref{subsec:anomalies}).



\section{Related Work}
\label{sec:related}

Multi-aspect evaluation measures for \ac{IR} have been studied for different tasks and aspects, starting from the \acs{INEX} initiative with relevance and coverage~\cite{KazaiEtAl2004}. 
Since then, measures have been proposed
to evaluate relevance and novelty or diversity, such as \acs{anDCG} \cite{ClarkeEtAl2008}, \acs{MAP-IA} \cite{AgrawalEtAl2009} and \acs{IA-ERR} \cite{ChapelleEtAl2009}; relevance, novelty and the amount of user effort, such as \acs{nCT} \cite{TangEtAl2017}; relevance, redundancy and user effort, such as \acs{RBU} \cite{AmigoEtAl2018}; relevance and understandability, such as \acs{uRBP} \cite{Zuccon2016} and the \ac{MM} framework \cite{PalottiEtAl2018}; and relevance and credibility, such as
\acs{NLRE}, \acs{NGRE}, \acs{nWCS}, \ac{CAM} and \acs{WHAM}
\cite{LiomaEtAl2017}. 
All these measures have limitations; we describe these next.


Firstly, except for \acs{RBU}, none of the above measures are based on a formal framework. They are defined as stand-alone tools to assess the effectiveness of a ranked list of documents. This means that, even if the measure can assess the effectiveness of an input ranking, the order induced by the measure over the space of input rankings is not well-defined. Hence, there is no canonical \emph{ideal ranking}\footnote{An \emph{ideal ranking} is the best ranking of all assessed documents for a given topic~\citep{JarvelinKekalainen2002}.}
that is well-defined or easy to compute, e.g., for \acs{anDCG}, the computation of the ideal ranking is equivalent to a minimal vertex covering problem~\cite{ClarkeEtAl2008}, an NP-complete problem, while for CT and \acs{nCT}, computing the ideal ranking is equivalent to the minimum edge dominating set problem~\cite{TangEtAl2017}, an NP-hard problem. 
Computationally better ways of comparing to an ideal ranking can be devised using graded similarity---so-called \emph{effectiveness levels} to an ideal ranking using Rank-Biased Overlap \citep{DBLP:conf/ictir/ClarkeVS20,DBLP:conf/cikm/ClarkeSV20,ClarkeEtAl2021}, but this approach requires defining a (set of) ideal ranking(s), which has not appeared for multi-aspect ranking prior to the present paper.

Evaluation measures that do not compare against an ideal ranking may be harder to interpret or problematic. 
\acs{DCG} is not upper bounded by $1$, thus different topics are not weighted equally and scores are not comparable.
Failing to compare against the ideal ranking is problematic in multi-aspect evaluation: $\alpha$-\acs{nDCG} allows systems to reach scores greater than $1$, which is supposed to be the score of the perfect system. With \acs{NLRE} and \acs{NGRE}, a system that retrieves no relevant or credible documents has error =$0$, i.e., achieves the best score, because the relative order of pairs of documents is always correct~\cite{LiomaEtAl2019}. Similarly, \acs{nWCS} can reach the perfect score of $1$, even if no relevant or credible documents are retrieved, since the normalization is computed with a re-ranking of the input ranking, instead of the ideal ranking.

Both \acs{uRBP} and \acs{RBU} have a different problem: to reach the perfect score of $1$, a system must retrieve an infinite number of relevant and understandable documents, even if those documents are not available in the collection. 
\ac{CAM} and \acs{WHAM} use the weighted arithmetic and weighted harmonic mean of any \ac{IR} measure computed with respect to relevance and  credibility independently. Therefore, depending on the distribution of labels across the aspects, it can be impossible for any system to reach the perfect score (see $\S$~\ref{subsec:anomalies}). 

Secondly, most of the above multi-aspect evaluation measures are defined for specific contexts and with a limited set of aspects, e.g.,~novelty, diversity, credibility and understandability, thus they cannot deal with a more general scenario and a variable number of aspects. For \acs{RBU}, even though a formal framework is defined, its formulation specifies only diversity and redundancy constraints, which cannot be applied to a general set of aspects.
This inability to generalise to more/other types of aspects means that, if a system must be evaluated with respect to a new aspect, the measure needs to be properly adapted. This can be easily done for some measures, e.g., \ac{CAM}, \acs{WHAM}, and \acs{nWCS}, but the lack of a formal framework behind them may lead to odd results, e.g., extending \acs{NLRE} to $3$ aspects returns a score distribution compressed towards $0$, preventing the rankings to be evaluated in a fair way~\cite{LiomaEtAl2019}.

\begin{figure*}[tb]
\begin{subfigure}{.18\linewidth}
\includegraphics[width=\textwidth]{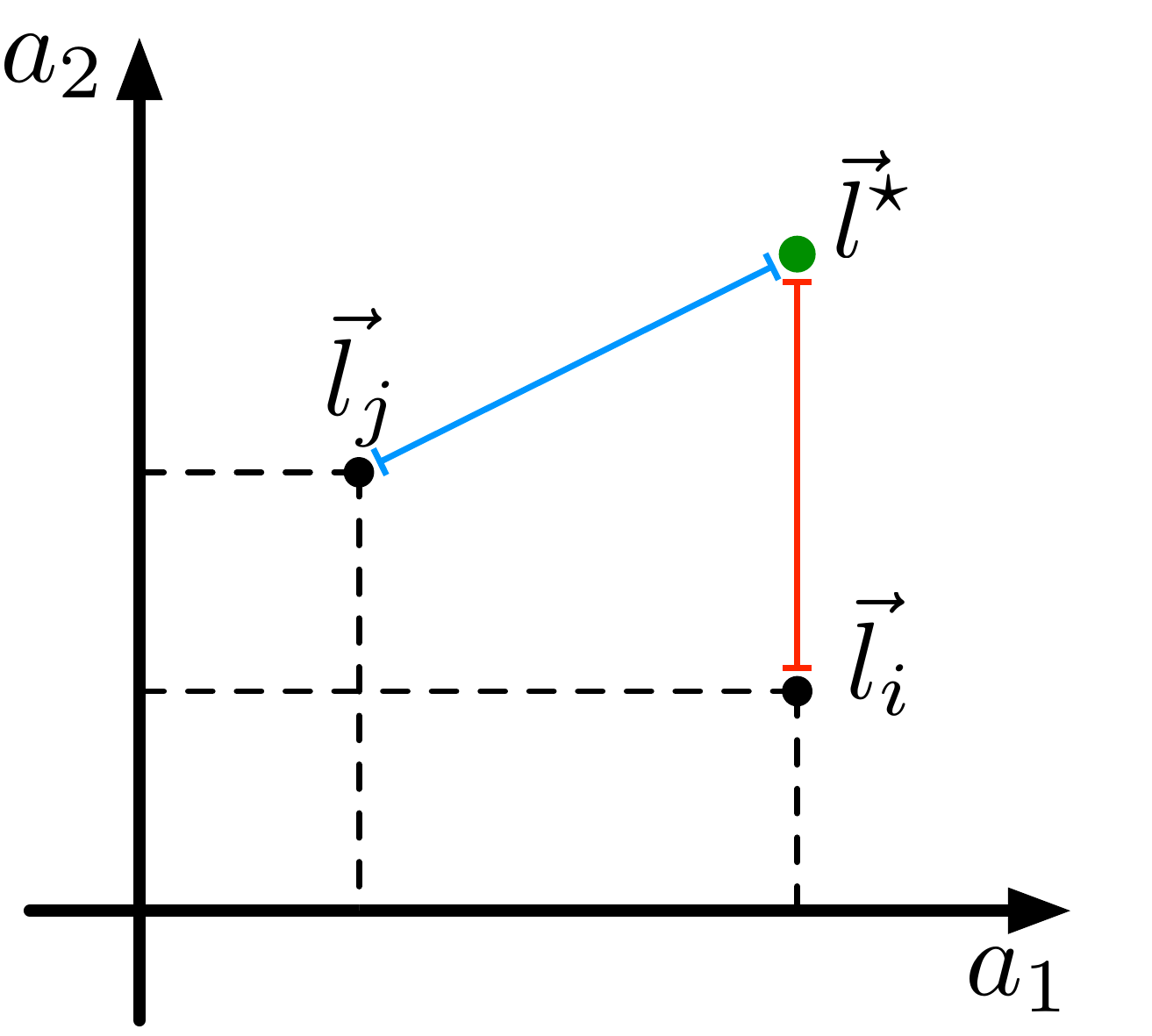}
\caption{Distance order.}
\label{fig:document_space}
\end{subfigure}%
\hfill
\begin{subfigure}{.17\linewidth}
\includegraphics[width=\textwidth]{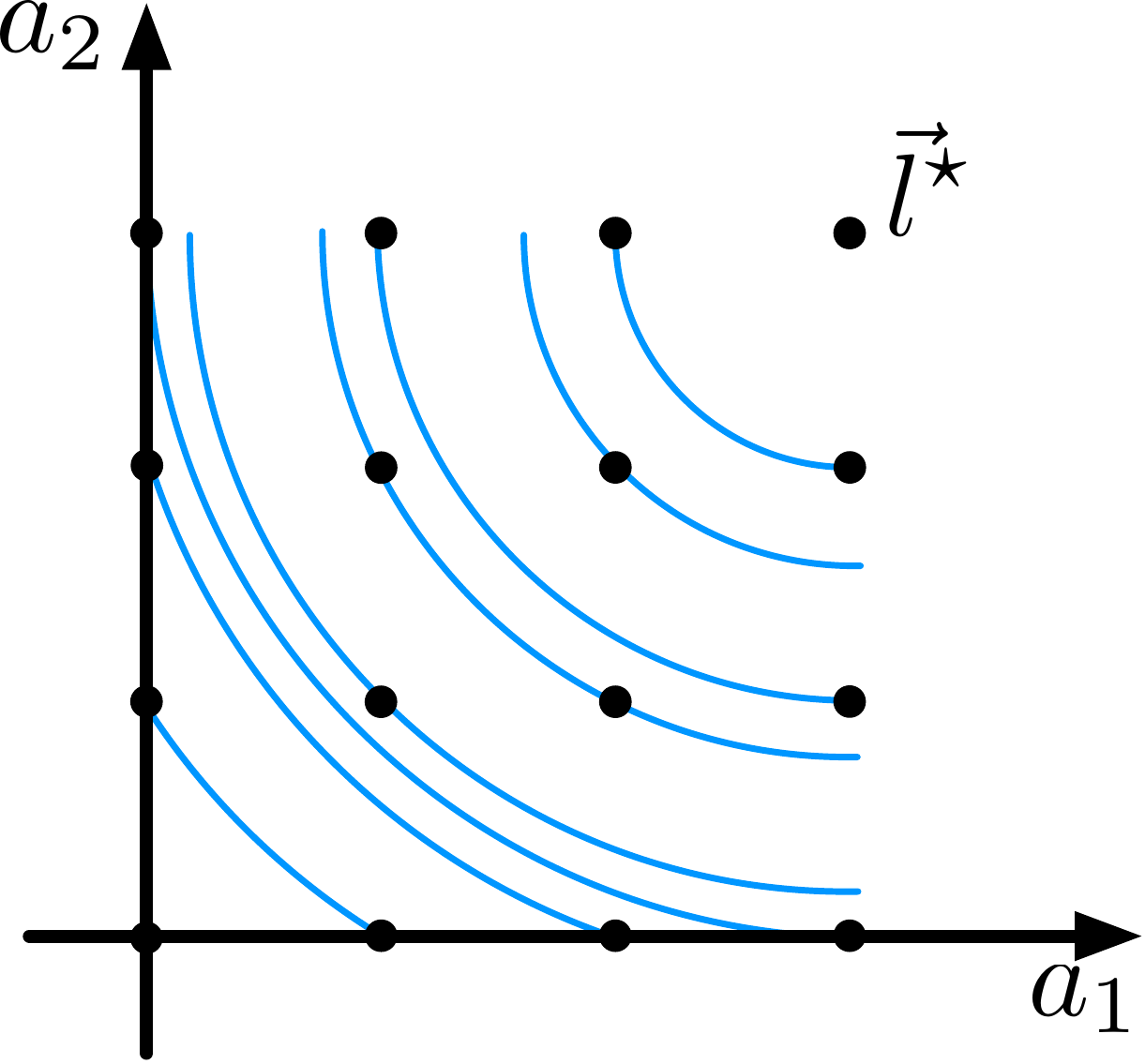}
\caption{Euclidean distance.}
\label{fig:dist_euclidean}
\end{subfigure}%
\hfill
\begin{subfigure}{.17\linewidth}
\centering
\includegraphics[width=\textwidth]{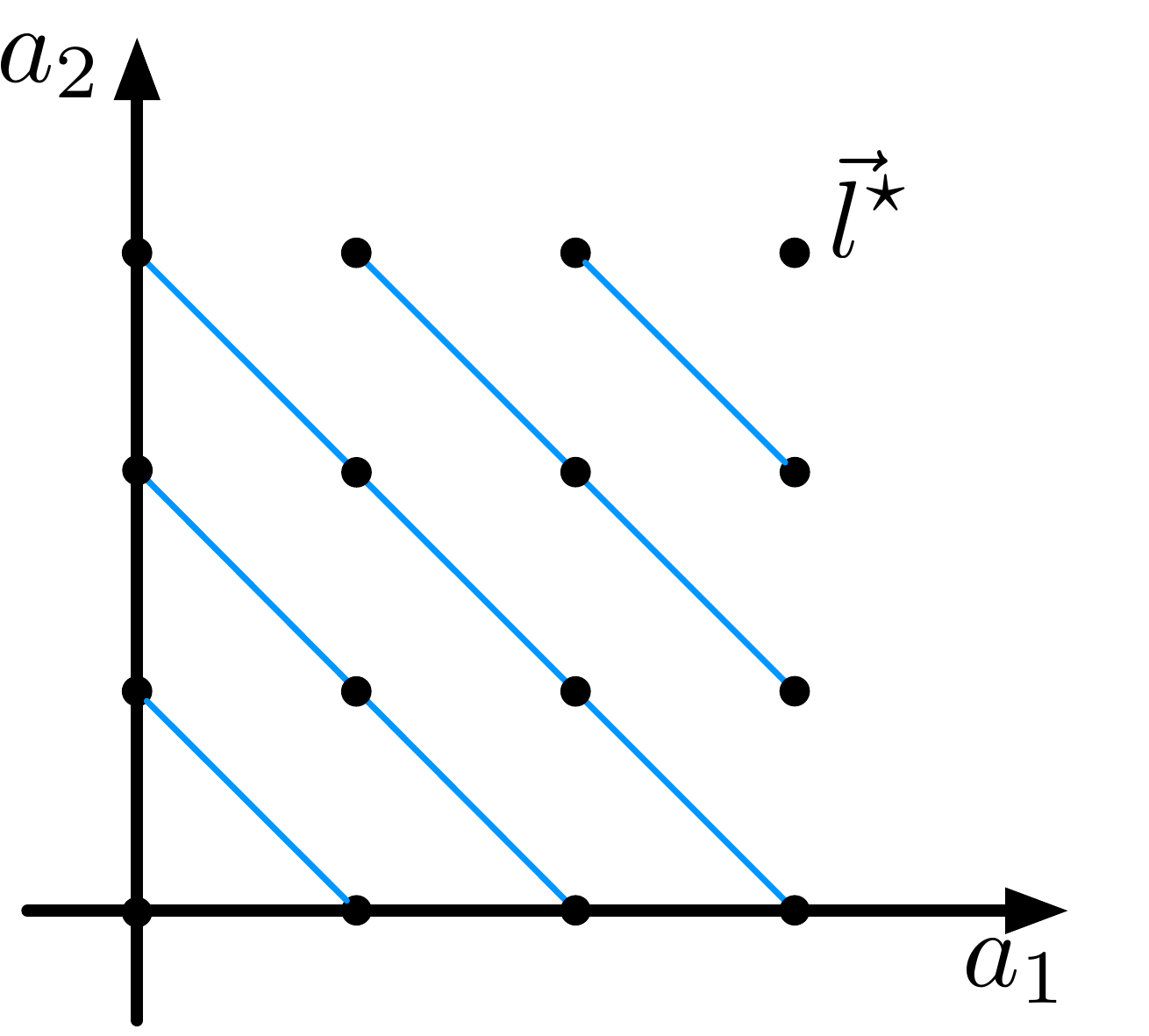}
\caption{Manhattan distance.}
\label{fig:dist_manhattan}
\end{subfigure}%
\hfill
\begin{subfigure}{.17\linewidth}
\centering
\includegraphics[width=\textwidth]{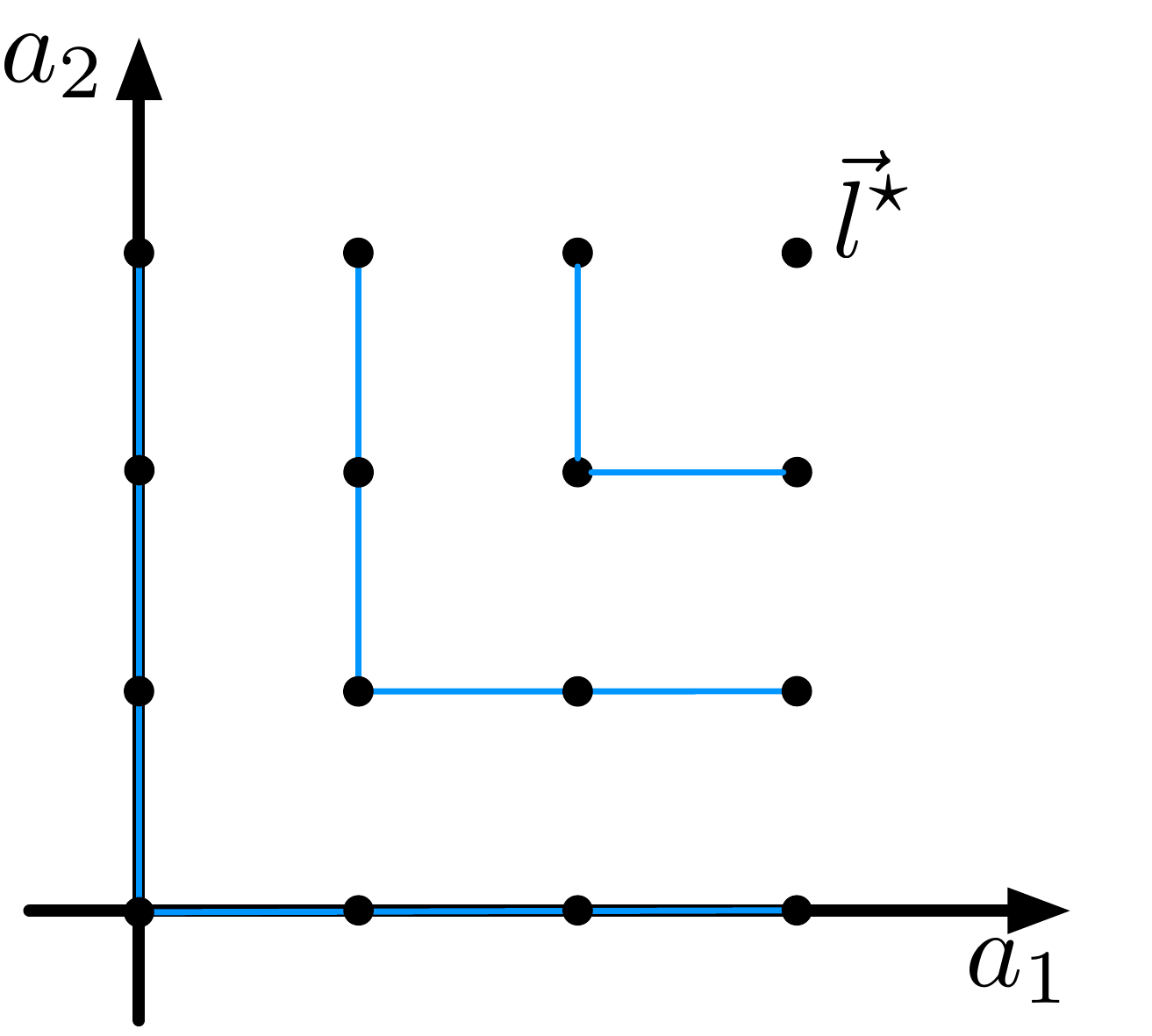}
 \caption{Chebyshev distance.}
\label{fig:dist_chebyshev}
\end{subfigure}%
\caption{Example with two aspects $a_1$ and $a_2$. Each point is a tuple of labels. The best label $\mathbf{l^\star}$ is in the top right. The distance between tuples of labels and $\mathbf{l^\star}$ defines a weak order relation. Blue lines connect tuples of labels at the same distance from $\mathbf{l^\star}$.}
\label{fig:dist_types}
\end{figure*}

\section{TOMA Framework}
\label{sec:approach}

We formalize the problem and our proposed methodology: we explain why reasoning in terms of multiple aspects leads to a partial order relation among documents ($\S$~\ref{subsec:problem}); how we complete the partial order relation with the distance order ($\S$~\ref{subsec:distance_order}); and how to use the distance order with state-of-the-art \ac{IR} evaluation measures ($\S$~\ref{subsec:integration_ir}).

\subsection{Formalization of the Problem}
\label{subsec:problem}

Let $A = \{a_1,\ldots,a_n\}$ be a set of \emph{aspects};
each aspect $a \in A$ has a non-empty set of \emph{labels} $L_{a} = \{l_{0}^{a},\ldots,l_{K_{a}}^{a}\}$ and an order relation $\prec_{a}$ such that:
$l_{0}^{a} \prec_{a} l_{1}^{a} \prec_{a} \cdots \prec_{a} l_{K_{a}}^{a}$, e.g., we may have $2$ aspects $A = \{\text{relevance},\text{correctness}\}$,
with the set $L_{r} = \{\texttt{nr}, \texttt{mr}, \texttt{fr}, \texttt{hr}\}$ (non-relevant, marginally relevant, fairly relevant, highly relevant) ordered as:
$\texttt{nr} \prec_{r} \texttt{mr} \prec_{r} \texttt{fr} \prec_{r} \texttt{hr}$;
and the set $L_{c} = \{\texttt{nc}, \texttt{pc}, \texttt{c}\}$ (non-correct, partially correct, correct)
ordered as:
$\texttt{nc} \prec_{c} \texttt{pc} \prec_{c} \texttt{c}$. Let $D$ be the set of \emph{documents} and $\topicset$ the set of \emph{topics}. 
Each document $d \in D$ is mapped to a \emph{ground truth} vector 
$\groundtruthmap(d, t) = (l_1,\ldots,l_n) \in L_{a_1} \times \cdots \times L_{a_n}$ that contains the ``true'' label of $d$ for each aspect, e.g., a document may have $\groundtruthmap(d, t) = (\textrm{highly relevant}, \textrm{non-correct})$.

In \ac{IR}, given a topic $t$, the objective is to 
rank documents in $D$ such that
for the documents $d^\prime, d \in D$, if $d^\prime$ is ranked before
$d$, then $\groundtruthmap(d, t) \preceq_{*} \groundtruthmap(d^\prime, t)$ for a given order relation $\preceq_{*}$. 
When there is only one aspect $A = \{a\}$, one can use $\prec_a$, the order on the set of labels $L_a$, to induce a weak order on $D$ and decide if $d^\prime$ should be ranked before $d$. If only relevance is assessed, we consider the relation induced by relevance labels, i.e., documents labelled ``highly relevant'' should be ranked before ``fairly/marginally relevant'' and ``non-relevant'' documents. Applying this approach to multiple aspects requires reasoning about orderings of tuples of labels with different aspects,  
e.g., for documents $d^\prime, d \in D$, such that $\groundtruthmap(d^\prime, t) = (\text{highly relevant}, \text{correct})$ and $\groundtruthmap(d, t) = ($marginally relevant, cor\-rect$)$, it is reasonable to rank $d^\prime$ before $d$. 

Indeed, there is one \emph{unequivocal} way of deeming one document better than another, and this is if document $d^\prime$ has better labels than document $d$ for \emph{every} aspect: if for
$\groundtruthmap(d, t) = (l_1,\ldots,l_n)$ and
$\groundtruthmap(d^\prime, t) = (l_1^\prime,\ldots,l_n^\prime)$ we have $l_i \preceq_{a_{i}} l_i^\prime$ for all $i \in \{1,\ldots,n\}$, then any document labeled $(l_1^\prime,\ldots,l_n^\prime)$
is better or equal than any document labelled $(l_1,\ldots,l_n)$ and should occur before it in a ``good'' ranking. We denote this order relation by $\groundtruthmap(d, t) \allpreceq \groundtruthmap(d^\prime, t)$.

The order relation $\allpreceq$ leads to a \emph{partial} instead of a \emph{total} order, i.e., there are documents that are \emph{not comparable}\footnote{A partial order is reflexive, antisymmetric and transitive; a total order is a partial order where all items are comparable; a weak order is a total order without antisymmetry~\cite{Halmos1974}.}, e.g.,
if $d^\prime$ is now highly relevant and partially correct, the final ranking is not clear: should one promote $d^\prime$ (more relevant) or $d$ (more correct)? 
This is an example of documents that are not comparable, so we have $\groundtruthmap(d, t) \not\allpreceq \groundtruthmap(d^\prime, t)$ and $\groundtruthmap(d^\prime, t) \not\allpreceq \groundtruthmap(d, t)$, and the choice of whether
$d^\prime$ is preferred to $d$ may lie on the intended application.

A partial order relation and the presence of not comparable documents imply that it is not possible to univocally rank the documents in $D$. 
If we could ``complete'' the partial order with a total order, or at least a weak order, we could rank documents and define an ideal ranking, where
for any $d^\prime, d \in D$, the order relation determines the rank position of $d^\prime$ and $d$. 
So, before tackling the problem of evaluating a ranked list of documents in a multi-aspect way, we build such an order relation.
This is detailed next.

\subsection{The distance order}
\label{subsec:distance_order}

We now explain how to obtain a weak order relation from the partial order relation $\allpreceq$. Consider the Cartesian product of all sets of labels $L = L_{a_1} \times \cdots \times L_{a_n}$. An element $\mathbf{l} \in L$ is a tuple of labels $\mathbf{l} = (l_1, \dots, l_n)$. The total order relation will be denoted by $\preceq_{*}$ and it will be a weak order relation on $L$, i.e., a total binary relation that is reflexive and transitive, but not necessarily anti-symmetric~\cite{FerranteEtAl2015b,FerranteEtAl2017,FerranteEtAl2018b}. This weak order allows \emph{all} tuples of labels to be compared, i.e., for any two
$\mathbf{l},\mathbf{l}^\prime \in L$ we will have
$\mathbf{l}^\prime \preceq_* \mathbf{l}$
and/or $\mathbf{l} \preceq_* \mathbf{l}^\prime$. Consequently, all documents will be comparable through their tuple of labels.

We require that the weak order relation $\preceq_{*}$ respects the partial order relation $\allpreceq$:
\begin{equation}
\label{eq:requirement_partial_order}
\forall \ \mathbf{l}, \mathbf{l}^\prime \in L \textrm{ we have } \mathbf{l} \allpreceq \mathbf{l}^\prime \Rightarrow \mathbf{l} \preceq_{*} \mathbf{l}^\prime
\end{equation}
This means that, for comparable documents, the partial order relation and the weak order relation rank documents in the same way. Moreover the weak order relation allows to rank even those documents that are not comparable with the partial order relation.

To define $\preceq_{*}$, we embed the tuples of labels in the Euclidean space and derive the weak order $\preceq_*$ using known distance functions. 
Let $g$ be an embedding function that maps tuples of labels in  Euclidean space $\mathcal{L} = \mathbb{R}^n$: $g(\mathbf{l}) = g(l_1, \ldots, l_n) = (g_{a_{1}}(l_1), \ldots, g_{a_{n}}(l_{n}))$. 
We assume that for each $a \in A$, $g_{a}$ is a non-decreasing map, i.e., for any $l, l^\prime \in L_{a}$ if $l \preceq_{a} l^\prime$ then $g_{a}(l) \leq g_{a}(l^\prime)$. 
Intuitively, $g_{a}$ assigns a number to each label, which allows to represent tuples of labels in the Euclidean space.
We illustrate in $\S$\ref{subsec:example} how the embedding function $g$ affects the final ranking of documents.

Through the embedding function $g$, each tuple of labels $\mathbf{l}$ is represented by a point in the Euclidean space $\mathcal{L}$ denoted by $\vec{l} = g(\mathbf{l})$. 
We define the \emph{best label tuple} as the tuple of labels $\mathbf{l^\star}$ whose coordinates are the best label for each aspect, $\mathbf{l^\star} = (l_{K_{a_1}}, \ldots, l_{K_{a_n}})$. 
The idea is to treat $\mathbf{l^\star}$ as
the maximum element and use the distance from this maximum element to define the desired weak order relation; e.g., for two aspects $a_1$ and $a_2$, each tuple of labels is represented as a point in the Euclidean plane, and the best label $\mathbf{l^\star}$ is represented by the topmost and right-most point (see Fig.~\ref{fig:document_space}). Then, given two documents $d$ and $d^\prime$, $d$ is ranked before $d^\prime$ if $\groundtruthmap(d, t)$ is closer to the best label than $\groundtruthmap(d^\prime, t)$. 

We formally define the \emph{distance order} as the following relation:
\begin{equation}
\label{eq:order_relation}
\mathbf{l} \preceq_{*} \mathbf{l^\prime} \ \iff \ \text{Dist}(\vec{l}, \vec{l^\star}) \geq \text{Dist}(\vec{l^\prime}, \vec{l^\star})
\end{equation}
where $\text{Dist} \colon \mathcal{L} \times \mathcal{L} \to [0, +\infty[$ is any function
such that $\text{Dist}(\vec{l}^\star,\vec{l}^\star) = 0$\footnote{Distance functions 
must be symmetric and satisfy the triangle inequality. Any such distance function satisfies our condition
on $\text{Dist}$, and so do our example distances.
}.
The relation $\preceq_{*}$ is a weak order: all $\mathbf{l}$, $\mathbf{l^\prime}$ are comparable because $\text{Dist}(\vec{l}, \vec{l^\star})$ is defined for all $\mathbf{l}$, and as $\geq$ is reflexive and transitive on $[0, +\infty[$, the relation $\preceq_{*}$ is reflexive and transitive (but not necessarily antisymmetric). 
%
%
Since the distance order is a weak order, it allows to deem items ``equally good'' when it is impossible or undesirable to impose a strict total order\footnote{This is the reason that \emph{weak} orders (that are not necessarily anti-symmetric), rather than strict total orders, are typically used in the literature \cite{DBLP:journals/jasis/Yao95,FerranteEtAl2018b}.}. Thus we write: 
\begin{equation}
\label{eq:equality_relation}
\mathbf{l} =_{*} \mathbf{l^\prime} \ \iff \ \text{Dist}(\vec{l}, \vec{l^\star}) = \text{Dist}(\vec{l^\prime}, \vec{l^\star})
\end{equation}
which means that $\vec{l}$ and $\vec{l^\prime}$ are at the same distance from $\vec{l}^{\star}$.

Note that the distance order can be \emph{tailored}: we may instantiate $\text{Dist}$ with any valid distance function. We illustrate this in Fig.~\ref{fig:dist_euclidean}-\ref{fig:dist_chebyshev} with \emph{Euclidean} 
(order relation $\preceq_{2}$), \emph{Manhattan} 
(order relation $\preceq_1$), and \emph{Chebyshev} 
(order relation $\preceq_\infty$). 
With these choices of $\text{Dist}$, the distance order defined in Eq.~\eqref{eq:order_relation}-\eqref{eq:equality_relation} respects the partial order $\allpreceq$, which means that it satisfies the requirement in Eq.~\eqref{eq:requirement_partial_order} because $g_a$ is a non decreasing map (a proof is provided in the online appendix\footnote{\url{https://github.com/lcschv/TOMA/blob/1d562036c50f7ff0a6df00246195098d7282b1ac/CIKM2021_appendix.pdf}}).

\subsection{Integration with \acs{IR} measures}
\label{subsec:integration_ir}

Next we integrate the distance order with known \ac{IR} measures such as AP or \acs{nDCG}. 
The binary relation $=_*$ in Eq.~\eqref{eq:equality_relation} is an equivalence relation. Given a tuple of labels $\mathbf{l}\in L$, its equivalence class $[\mathbf{l}]_{*}$ is the set of all tuples of labels with equal distance from the best label $[\mathbf{l}]_{*} = \{\mathbf{l^\prime}\in L \colon \text{Dist}(\vec{l^\prime}, \vec{l^\star}) = \text{Dist}(\vec{l}, \vec{l^\star})\}$. 

Inducing the relation defined in Eq.~\eqref{eq:order_relation} on the set of documents $D$ allows to rank documents by their membership to each equivalence class, which corresponds to the distance of their tuple of labels to the best label. We place closest to the top of the ranking documents whose equivalence class is closest to the best label, and vice versa.

To combine the distance order with \ac{IR} measures we map each equivalence class (set of tuple of labels), to a non negative integer. This is similar to what happens with single-aspect evaluation, where each label is mapped to a weight: e.g., with $4$ relevance labels, we can compute \acs{nDCG} with equi-spaced relevance weights $\{0, 1, 2, 3\}$~\cite{JarvelinKekalainen2002} or exponential weights $\{0, 2, 4, 8\}$~\cite{BurgesEtAl2005}. 
We define a \emph{weight function} $W \colon L \to \mathbb{N}_{0}^+$ as a map such that the order relation $\preceq_{*}$ is preserved:
\begin{equation}
\label{eq:weighted_function}
    \forall \ \mathbf{l}, \mathbf{l^\prime} \in L \colon \mathbf{l} \preceq_{*} \mathbf{l^\prime} \implies W(\mathbf{l}) \leq W(\mathbf{l^\prime})
\end{equation}
where the constraint in Eq.~\eqref{eq:weighted_function} entails that $W$ is a non-decreasing function with respect to the weak order $\preceq_{*}$ on the set of tuples of labels. This means that $W$ can return different integers for each equivalence class, but also the same integer for different equivalence classes, i.e., $0$ and $1$, whenever we need to compute a binary single-aspect \ac{IR} measure as \acs{AP}.

To summarize, our \ac{TOMA} method has $3$ steps:
\begin{enumerate}
\item We embed tuples of labels into elements of Euclidean space, and we derive the weak order $\preceq_*$ using a distance function;
\item We define an adjustable weight function $W$ that preserves $\preceq_{*}$ and maps each tuple of labels to a single integer weight (this allows to aggregate tuple of labels so that better documents can be given greater weight);
\item Having such a weak order and the weight function $W$, any existing single-aspect \ac{IR} evaluation measure can be used to assess the quality. Thus, we choose a single-aspect evaluation measure $\mu$ and compute the final evaluation score as $
M=\mu \circ W\colon  M(r_{t}) = \mu(W(\textrm{GT}(d_1, t)), \ldots, W(\textrm{GT}(d_N, t)))$, where $r_{t}$ is a ranked list of documents.
\end{enumerate}
The above is compatible with any number and type of aspect.

\subsection{Example}
\label{subsec:example}

\begin{table*}[tb]
\caption{Final ordering of tuples of labels embedded in the Euclidean space. Relevance labels are always embedded in the same mapping (under \textit{Relevance}). We use different mappings for correctness labels (under \textit{Correctness}). Tuples that are relevant and not correct (high-traffic fake news) are in \textcolor{notcorrect}{red}.}
\label{tab:example}
\begin{tabular}{@{}llll@{}}
\toprule
Relevance               & Correctness            & Distance  & Order among Tuples of Labels               \\ \midrule
\multirow{3}{*}{$\{0, 1, 2, 3\}$} & \multirow{3}{*}{$\{0, 3/2, 3\}$} & Euclidean & $(3,3)\preceq_{*}(2,3)\preceq_{*}(3, 3/2)\preceq_{*}(2,3/2)\preceq_{*}(1,3)\preceq_{*}(1,3/2)\preceq_{*}\mathbin{\textcolor{notcorrect}{(3,0)}}\preceq_{*}\mathbin{\textcolor{notcorrect}{(2,0)}}\preceq_{*}\mathbin{\textcolor{notcorrect}{(1,0)}}\preceq_{*}(0,0)$ \\
                                  &                                  & Manhattan &  $(3,3)\preceq_{*}(2,3)\preceq_{*}(3, 3/2)\preceq_{*}(1,3)\preceq_{*}(2,3/2)\preceq_{*}\mathbin{\textcolor{notcorrect}{(3,0)}}\preceq_{*}(1,3/2)\preceq_{*}\mathbin{\textcolor{notcorrect}{(2,0)}}\preceq_{*}\mathbin{\textcolor{notcorrect}{(1,0)}}\preceq_{*}(0,0)$   \\
                                  &                                  & Chebyshev & $(3,3)\preceq_{*}(2,3)\preceq_{*}(3, 3/2)=_{*}(2,3/2)\preceq_{*}(1,3)=_{*}(1,3/2)\preceq_{*}\mathbin{\textcolor{notcorrect}{(3,0)}}=_{*}\mathbin{\textcolor{notcorrect}{(2,0)}}=_{*}\mathbin{\textcolor{notcorrect}{(1,0)}}=_{*}(0,0)$ \\
\midrule
\multirow{3}{*}{$\{0, 1, 2, 3\}$} & \multirow{3}{*}{$\{0, 1, 2\}$}   & Euclidean &  $(3,2)\preceq_{*}(3,1)=_{*}(2,2)\preceq_{*}(2,1)\preceq_{*}\mathbin{\textcolor{notcorrect}{(3,0)}}=_{*}(1,2)\preceq_{*}\mathbin{\textcolor{notcorrect}{(2,0)}}=_{*}(1,1)\preceq_{*}\mathbin{\textcolor{notcorrect}{(1,0)}}\preceq_{*}(0,0)$ \\
                                  &                                  & Manhattan & $(3,2)\preceq_{*}(3,1)=_{*}(2,2)\preceq_{*}\mathbin{\textcolor{notcorrect}{(3,0)}}=_{*}(2,1)=_{*}(1,2)\preceq_{*}\mathbin{\textcolor{notcorrect}{(2,0)}}=_{*}(1,1)\preceq_{*}\mathbin{\textcolor{notcorrect}{(1,0)}}\preceq_{*}(0,0)$ \\
                                  &                                  & Chebyshev & $(3,2)\preceq_{*}(3,1)=_{*}(2,1)=_{*}(2,2)\preceq_{*}\mathbin{\textcolor{notcorrect}{(3,0)}}=_{*}\mathbin{\textcolor{notcorrect}{(2,0)}}=_{*}\mathbin{\textcolor{notcorrect}{(1,0)}}=_{*}(1,1)=_{*}(1,2)\preceq_{*}(0,0)$ \\
\midrule
\multirow{3}{*}{$\{0, 1, 2, 3\}$} & \multirow{3}{*}{$\{0, 2, 6\}$}  & Euclidean & $(3,6)\preceq_{*}(2,6)\preceq_{*}(1,6)\preceq_{*}(3,2)\preceq_{*}(2,2)\preceq_{*}(1,2)\preceq_{*}\mathbin{\textcolor{notcorrect}{(3,0)}}\preceq_{*}\mathbin{\textcolor{notcorrect}{(2,0)}}\preceq_{*}\mathbin{\textcolor{notcorrect}{(1,0)}}\preceq_{*}(0,0)$ \\
                                  &                                  & Manhattan & $(3,6)\preceq_{*}(2,6)\preceq_{*}(1,6)\preceq_{*}(3,2)\preceq_{*}(2,2)\preceq_{*}(1,2)=_{*}\mathbin{\textcolor{notcorrect}{(3,0)}}\preceq_{*}\mathbin{\textcolor{notcorrect}{(2,0)}}\preceq_{*}\mathbin{\textcolor{notcorrect}{(1,0)}}\preceq_{*}(0,0)$ \\
                                  &                                  & Chebyshev & $(3,6)\preceq_{*}(2,6)\preceq_{*}(1,6)\preceq_{*}(3,2)=_{*}(2,2)=_{*}(1,2)\preceq_{*}\mathbin{\textcolor{notcorrect}{(3,0)}}=_{*}\mathbin{\textcolor{notcorrect}{(2,0)}}=_{*}\mathbin{\textcolor{notcorrect}{(1,0)}}=_{*}(0,0)$ \\
\bottomrule
\end{tabular}%
\end{table*}
We present an example on the role of different choices of embedding, distance, and weight functions in \ac{TOMA} with $4$ relevance labels $\{\texttt{nr},\texttt{mr},\texttt{fr},\texttt{hr}\}$ and $3$ correctness labels $\{\texttt{nc}, \texttt{pc}, \texttt{c}\}$. As in real scenarios~\cite{LiomaEtAl2019}, we assume that not relevant documents are not correct: as they do not include information about the topic, they cannot be correct with respect to that topic.

Tab.~\ref{tab:example} shows $3$ different embeddings for correctness; the embedding for relevance is fixed. 
Note that the distance functions are invariant under translations and rotations, thus, rather than the actual values assigned from the embedding function $g$, it is important to consider the relation between different aspects. 
Independently of the choice of the embedding function and due to the definition of the selected distance functions, we see that: (i) Chebyshev generates the least number of equivalence classes and deems many tuples of labels as equal, since by taking the maximum it considers just the ``furthest'' or worst aspect to compute the distance; (ii) Manhatthan is somehow in-between Chebyshev and Euclidean and generates the equivalence classes by taking the sum across aspects; (iii) Euclidean generates the highest number of equivalence classes as it differentiates among tuples more than Manhatthan and is more sensitive to extreme cases, e.g., cases where one aspect has the best label and all other aspects have the lowest label. 

In the 1$^{st}$ scenario of Tab.~\ref{tab:example} we map relevance and correctness to the same interval $[0, 3]$ (i.e., a highly relevant document is as ``important'' as a correct document). All labels are equi-spaced in the given range (the difference between a fairly relevant and a marginally relevant document is the same as that between a highly relevant and a fairly relevant one). 
With the Euclidean distance all relevant and not correct documents will be deemed worse than all other documents, but will be placed before not relevant and not correct documents. 
On the other hand, Chebyshev places relevant and not correct documents in the same equivalence class as not relevant documents, so those documents do not provide any contribution and can be simply filtered out. Manhattan is a middle solution: highly relevant and not correct documents are deemed better than marginally relevant and partially correct documents, but worse than all other correct or partially correct documents. 

In the 2$^{nd}$ scenario of Tab.~\ref{tab:example} relevance and correctness are mapped to different ranges, but all labels are equi-spaced with the same step of size $1$. Here, relevance is more important than correctness. This is reflected on the sorting of equivalence classes: for all distance functions, highly relevant and not correct documents do not belong to the worst equivalence classes, but they are somehow better than partially correct documents. Even Chebyshev, which can be seen as the ``strictest'' distance function, places all relevant and not correct documents in the same class, which is considered better than the class of not relevant and not correct documents.

In the 3$^{rd}$ scenario of Tab.~\ref{tab:example} correctness is mapped to a range twice the size as the relevance range and we do not use equi-spaced labels for correctness. We assign more importance to correctness than relevance, and among correctness labels we penalize not correct and partially correct documents. The result is that for all distance functions relevant and not correct documents are considered among the worst equivalence classes. This particular setting affects also the other equivalence classes: correctness is preferred over relevance, e.g., correct documents should be always ranked before partially correct documents, regardless of their relevance label.

Note that \ac{TOMA} requires a weight function satisfying the requirement in Eq.~\eqref{eq:weighted_function}. If we wish to reward a system for sorting documents exactly as presented by the equivalence classes in Tab.~\ref{tab:example}, then the weight function should assign a different integer to each equivalence class. This choice of weight is similar to the choice of weights for relevance labels and its impact on the evaluation outcome is strictly tight to the evaluation measure used, as for example when one considers \ac{nDCG} with different weighting schemes~\cite{JarvelinKekalainen2002}.

\begin{table*}[tb]
\caption{Experimental data. All aspects are labelled by TREC except popularity ($\dagger$ approximated by PageRank) and non-spamminess ($\ddagger$ approximated by Waterloo Spam Ranking). * means that the junk labels are merged with non relevant.}
\label{tab:data}
\resizebox{\textwidth}{!}{%
\begin{tabular}{l|c|c|c|c|c|c|c|c|c|c}
\toprule
 &\multicolumn{10}{c}{\textbf{TREC tracks}}\\
 &\textbf{Web 2009}&\textbf{Web 2010}&\textbf{Web 2011}&\textbf{Web 2012}&\textbf{Web 2013}&\textbf{Web 2014}&\textbf{Task 2015}&\textbf{Task 2016}&\textbf{Decision 2019}&\textbf{Misinfo2020}\\
\midrule
\textbf{Collection}&\multicolumn{4}{c|}{ClueWeb09} & \multicolumn{4}{c|}{ClueWeb12} & ClueWeb12-B13 &CommonCrawl News \\
\midrule
\textbf{Topics}&50&48&50&50&50&50& 35 & 50 & 50& 46 \\
\midrule
\textbf{Submitted runs}&71&56&61&48&61&30& 6 & 9 & 32 & 51\\
\midrule
&\multicolumn{1}{c|}{relevance (4)}&\multicolumn{1}{c|}{relevance (5*)}&\multicolumn{1}{c|}{relevance (4*)}&\multicolumn{3}{c|}{relevance (5*)}& \multicolumn{2}{c|}{relevance (3*)} & relevance (3) & relevance (2)  \\
\textbf{Aspects}& popularity$\dagger$ (3)&\multicolumn{1}{c|}{popularity$\dagger$ (3)}
&\multicolumn{1}{c|}{popularity$\dagger$ (3)}
&\multicolumn{3}{c|}{popularity$\dagger$ (3)}&\multicolumn{2}{c|}{usefulness (3)} & credibility (2) &credibility (2)\\
\textbf{(label grades)}&non-spam$\ddagger$ (3)& \multicolumn{1}{c|}{non-spam$\ddagger$ (3)}
& \multicolumn{1}{c|}{non-spam$\ddagger$ (3)}& \multicolumn{3}{c|}{non-spam$\ddagger$ (3)}& \multicolumn{2}{c|}{popularity$\dagger$ (3)} & correctness (2) &correctness (2) \\
&&&&\multicolumn{3}{c|}{ }& \multicolumn{2}{c|}{non-spam$\ddagger$ (3)} & \\
\bottomrule
\end{tabular}%
}
\end{table*}

\section{Experimental Evaluation}
\label{sec:ex_setup}
We evaluate \ac{TOMA} 
on $425$ rankings that were submitted as official runs to $10$ TREC tracks~\citep{DBLP:conf/trec/2016,DBLP:conf/trec/YilmazVMKCC15,LiomaEtAl2019,overview2009,overview2010,overview2011,overview2012,overview2013,overview2014,overview-misininfo-2020} (see Tab.~\ref{tab:data}). 

\subsection{Experimental Setup}
We use up to $5$ different aspects. All aspects are assessed by TREC assessors as part of the corresponding track, 
except \textit{popularity} and \textit{non-spamminess}. 
We approximate \textit{popularity} by PageRank\footnote{\url{http://www.lemurproject.org/clueweb12/PageRank.php}}, and \textit{non-spamminess} by the Waterloo Spam Ranking\footnote{\url{https://www.mansci.uwaterloo.ca/~msmucker/cw12spam/}}.
We discretize the PageRank scores to generate $3$ grades of popularity (not popular, fairly popular, highly popular), while simulating a power law distribution of popular and not popular documents ($5\%$ highly popular, $10\%$ fairly popular, and $85\%$ not popular). For non-spamminess, we generate $3$ grades of labels (spam, fairly spam, not spam) from the Waterloo Spam Ranking.
We treat any document with score $< 80$ as spam ($77\%$), documents with score in $[80, 89]$ ($14\%$) as fairly spam, and documents with score $\geq 90$ ($9\%$) as not spam~\cite{Petersen:2015:EGB:2808194.2809458}.

For the Web 2010-2014 and Task 2015-2016 tracks, we merge the labels junk and non relevant into non relevant, as was done by the TREC track organisers.  
For Task 2015-2016, Decision 2019 and Misinformation 2020, \emph{usefulness}, \emph{credibility}, and \emph{correctness} were not assessed for not relevant documents, thus not relevant documents are assumed to be not useful, not credible, and not correct.

We evaluate $3$ versions of our method \ac{TOMA}, with Euclidean, Manhattan, and Chebyshev, as per the distance metric used in Eq.~\eqref{eq:order_relation} (abbreviated as EUCL, MANH, and CHEB henceforth). 
We compare these to two state-of-the-art baselines, \ac{CAM}~\cite{LiomaEtAl2017} and \ac{MM}~\cite{PalottiEtAl2018}. 

Given a set of aspects $A$\footnote{\ac{CAM} was originally formulated for two aspects~\cite{LiomaEtAl2017}.}, \ac{CAM} aggregates their scores through a weighted average:
\begin{equation}
\label{eq:cam_ext}
    \text{CAM}(r_{t})=\sum_{a \in A}{p_{a}\times \mu(\hat{r}_{t,a})}
\end{equation}
where $\mu(\cdot)$ is the evaluation measure (e.g., \acs{nDCG}), $\hat{r}_{t,a}$ is the ranking labelled with respect to aspect $a$, and $p_{a}$ is a parameter controlling the importance of each aspect: $p_{a} \in [0,1]$ and $\sum_{a \in A} p_{a} = 1$.

\ac{MM}~\cite{PalottiEtAl2018} aggregates the evaluation measure scores computed for each aspect individually with a weighted harmonic mean:
\begin{equation}
\label{eq:mm_def}
\text{MM}(r_{t}) = \frac{\sum_{a \in A}p_{a}}{\sum \limits_{a \in A} \frac{p_{a}}{\mu(\hat{r}_{t,a})}}  
\end{equation}
with the same notation as above. 
 
Out of the other multi-aspect methods presented in $\S$\ref{sec:related}, we do not use \acs{WHAM}~\cite{LiomaEtAl2017} as baseline because it also uses the weighted harmonic mean to aggregate the evaluation measure scores. However, \acs{WHAM} is defined only for relevance and credibility, and can therefore be seen as an instantiation of \ac{MM} restricted to two aspects. All other multi aspect measures in $\S$\ref{sec:related} need a predefined set and number of aspects, thus are not applicable in our scenario.

We instantiate our method and the baselines using (1) \ac{nDCG}~\cite{KekalainenJarvelin2002} and graded labels (when available); and using (2) \ac{AP}~\cite{BuckleyVoorhees2005} and binary labels (we convert all graded labels to binary by treating all grades above zero as one, and grades equal/below zero as zero). We consider all aspects equally important (all aspects are mapped to an integer scale with one unit separating each grade). All source code is publicly available\footnote{\url{https://github.com/lcschv/TOMA.git}}.

\subsection{Anomalies of CAM \& MM}
\label{subsec:anomalies}

Next we discuss anomalies of CAM and MM that TOMA overcomes.

\begin{table*}[htb]
\caption{\ac{CAM}, \ac{MM} and \ac{TOMA} scores instantiated with \ac{AP} \& \ac{nDCG} for all rankings in $D$. The highest scores are in bold.}
\label{tab:AP_MM_CAM}
\centering
\resizebox{\textwidth}{!}{%
\begin{tabular}{@{}llllll||llllll||llllll@{}}
\toprule
\multicolumn{18}{c}{AP}\\ \midrule
Length $3$                & \ac{CAM} & \ac{MM}  & EUCL & MANH & CHEB & Length $2$         & \ac{CAM} & \ac{MM}  & EUCL & MANH & CHEB & Length $1$  & \ac{CAM} & \ac{MM}  & EUCL & MANH & CHEB \\ \midrule
$(d_{1}, d_{2}, d_{3})$ & $\mathbf{0.7917}$ & $\mathbf{0.3684}$ & $\mathbf{1}$ & $\mathbf{1}$ & $0.5$ & $(d_{1}, d_{2})$ & $0.6250$ & $0.25$ & $\mathbf{1}$ & $\mathbf{1}$ & $0.5$ & $(d_{1})$ & $0.5$    & $0$ & $0.5$ & $0.5$ & $0$     \\
$(d_{1}, d_{3}, d_{2})$ & $\mathbf{0.7917}$ & $\mathbf{0.3684}$ & $0.8333$ & $0.8333$ & $0.3333$ & $(d_{1}, d_{3})$ & $0.6250$ & $0.25$ & $0.5$ & $0.5$ & $0$ & $(d_{2})$ & $0.25$   & $0$ & $0.5$ & $0.5$ & $\mathbf{1}$     \\
$(d_{2}, d_{1}, d_{3})$ & $0.6667$ & $0.3125$ & $\mathbf{1}$ & $\mathbf{1}$ & $\mathbf{1}$ & $(d_{2}, d_{1})$ & $0.5$    & $0.25$  & $\mathbf{1}$ & $\mathbf{1}$ & $\mathbf{1}$ & $(d_{3})$ & $0.25$   & $0$     & $0$ & $0$ & $0$ \\
$(d_{2}, d_{3}, d_{1})$ & $0.6667$ & $0.25$ & $0.8333$ & $0.8333$ & $\mathbf{1}$   & $(d_{2}, d_{3})$ & $0.5$    & $0$ & $0.5$ & $0.5$ & $\mathbf{1}$     & -         & -        & -     & -         & -        & -  \\
$(d_{3}, d_{1}, d_{2})$ & $0.6667$ & $0.3125$ & $0.5833$ & $0.5833$ & $0.3333$ & $(d_{3}, d_{1})$ & $0.5$    & $0.25$  & $0.25$ & $0.25$ & $0$ & -         & -        & -  & -         & -        & -       \\
$(d_{3}, d_{2}, d_{1})$ & $0.6667$ & $0.25$  & $0.5833$ & $0.5833$ & $0.5$ & $(d_{3}, d_{2})$ & $0.5$    & $0$ & $0.25$ & $0.25$ & $0.5$ & -         & -        & -   & -         & -        & -       \\ \bottomrule
%
%
\toprule
\multicolumn{18}{c}{NDCG}\\ \midrule
Length $3$                & \ac{CAM} & \ac{MM}  & EUCL & MANH & CHEB & Length $2$         & \ac{CAM} & \ac{MM}  & EUCL & MANH & CHEB & Length $1$  & \ac{CAM} & \ac{MM}  & EUCL & MANH & CHEB \\ \midrule
$(d_{1}, d_{2}, d_{3})$ & $\mathbf{0.9073}$ & $0.4489$ & $0.9367$ & $0.9711$ & $0.8597$ & $(d_{1}, d_{2})$ & $0.7682$ & $0.3491$ & $0.8080$ & $0.8147$ & $0.8597$ & $(d_{1})$ & $0.4728$ & $0.1491$ & $0.4290$ & $0.4693$ & $0.3801$ \\
$(d_{1}, d_{3}, d_{2})$ & $0.8824$ & $0.4386$ & $0.8917$ & $0.9404$ & $0.7602$ & $(d_{1}, d_{3})$ & $0.6483$ & $0.3145$ & $0.5914$ & $0.6667$ & $0.3801$ & $(d_{2})$ & $0.4682$ & $0.2258$ & $0.6006$ & $0.5475$ & $0.7602$ \\
$(d_{2}, d_{1}, d_{3})$ & $0.9056$ & $\mathbf{0.4516}$ & $\mathbf{1}$ & $\mathbf{1}$ & $\mathbf{1}$ & $(d_{2}, d_{1})$ & $0.7665$ & $0.3776$ & $0.8713$ & $0.8436$ & $\mathbf{1}$ & $(d_{3})$ & $0.2781$ & $0$ & $0.2574$ & $0.3129$ & $0$ \\
$(d_{2}, d_{3}, d_{1})$ & $0.8801$ & $0.4319$ & $0.9775$ & $0.9795$ & $0.9502$ & $(d_{2}, d_{3})$ & $0.6437$ & $0.2679$ & $0.7630$ & $0.7449$ & $0.7602$ & -         & -        & -  & -         & -        & -  \\
$(d_{3}, d_{1}, d_{2})$ & $0.8106$ & $0.3930$ & $0.8284$ & $0.8827$ & $0.6199$ & $(d_{3}, d_{1})$ & $0.5765$ & $0.2801$ & $0.5281$ & $0.6089$ & $0.2398$ & -         & -        & -  & -         & -        & -  \\
$(d_{3}, d_{2}, d_{1})$ & $0.8100$ & $0.3827$ & $0.8509$ & $0.8929$ & $0.6697$ & $(d_{3}, d_{2})$ & $0.5735$ & $0.1897$ & $0.6364$ & $0.6583$ & $0.4796$ & -         & -        & -      & -         & -        & -       \\ \bottomrule
\end{tabular}%
}
\end{table*}

\begin{figure}[tb]
    \begin{subfigure}{.4\textwidth}
  \centering
  \includegraphics[width=\textwidth]{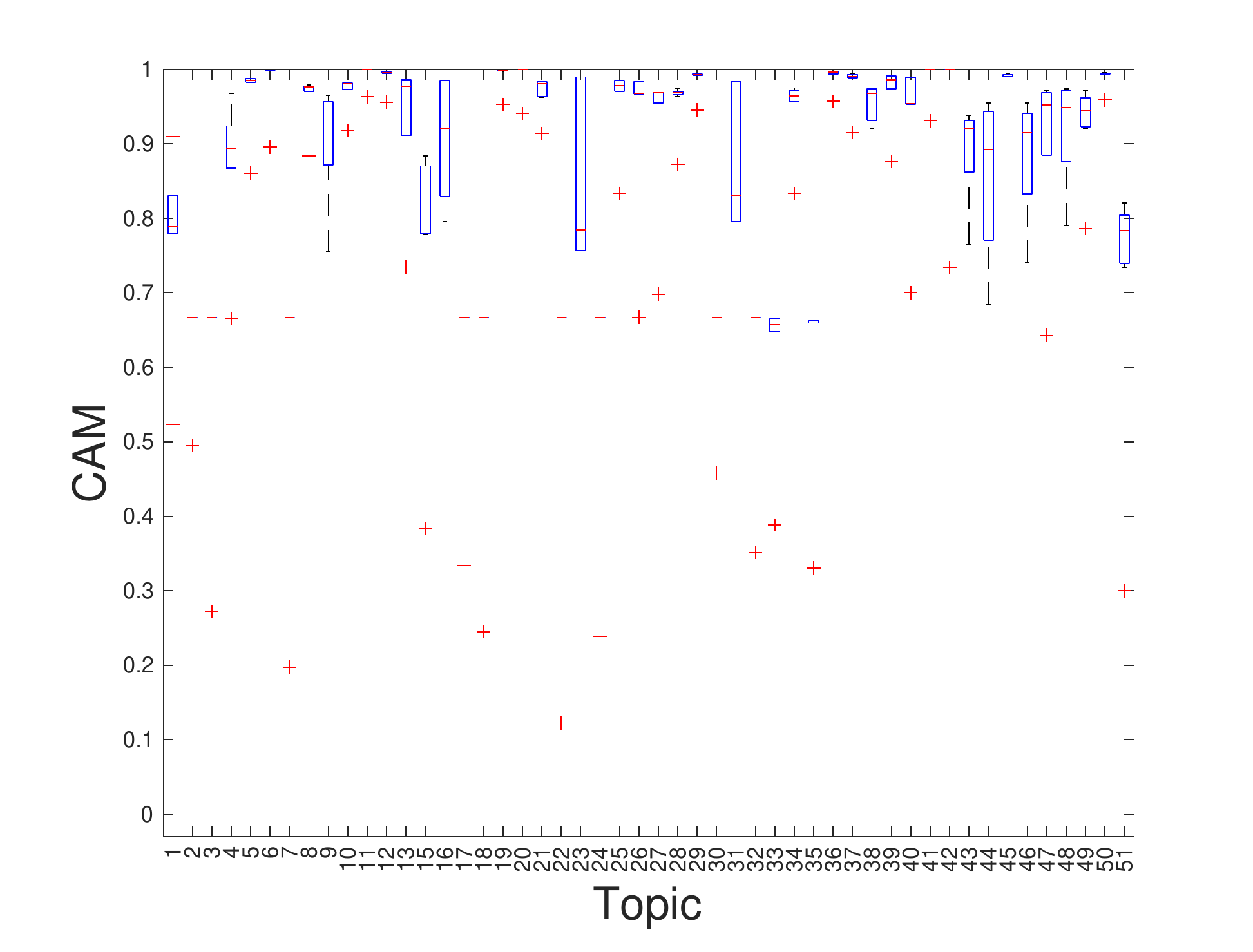}
  \end{subfigure}
  \begin{subfigure}{.4\textwidth}
  \centering
  \includegraphics[width=\textwidth]{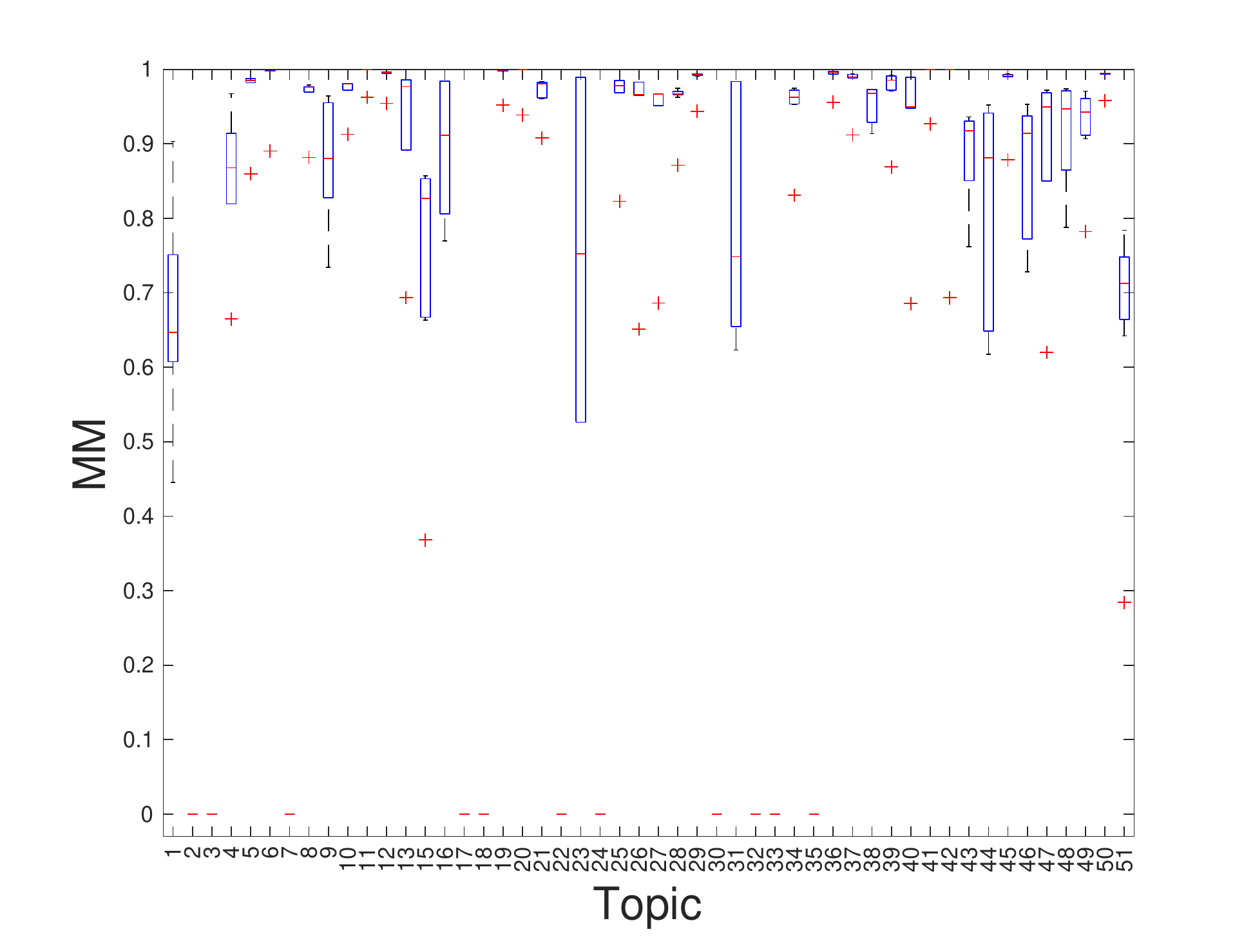}
  \end{subfigure}
  \caption{Box-plots for \ac{CAM} and \ac{MM} with \ac{nDCG} on the Decision Track 2019. Topic numbers are on the $x$-axis and measures scores on the $y$-axis. The maximum value for \ac{CAM} and \ac{MM} is variable and depends on the topic and the aspects.}
  \label{fig:plot_measures_ranges_cam}
\end{figure}

\paragraph{Problem 1: \ac{MM} is ill-defined.}
As the harmonic mean is not defined with zero values, \acs{MM} is not defined if $\exists a \in A$ such that $\mu(\hat{r}_{t,a}) = 0$, e.g.,~a ranking does not retrieve any correct or relevant document. 
To compute \acs{MM} even in these cases, as 
the denominator in Eq.~\eqref{eq:mm_def} tends to $+\infty$ if any
$\mu(\hat{r}_{t,a})$ tends to zero, we set $\text{MM}(r_{t}) = 0$.
For classification measures, this problem is called the Strong Definiteness Axiom~\cite{Sebastiani2015}.
It represents a serious issue for collections where there are a few documents with a positive label for certain aspects.
For example, for the Task Tracks, since useful documents are very sparse, many systems are not able to retrieve any useful document and they all have a $0$ score, independently of the number of relevant documents they retrieve.
\ac{TOMA} does not have this problem, because we first assign a weight to each tuple of labels and then compute a single-aspect evaluation measure $\mu$, thus there is no division by $0$ and \acs{TOMA} is well defined.

\paragraph{Problem 2: CAM and MM can range in different intervals.}
Given a set of documents $D$ and a set of aspects $A$, by definition \acs{CAM} and \acs{MM} are
multi-aspect evaluation measures $M \colon D^* \to 
[0, X]$, where $D^*$ is the set of rankings and $X \leq 1$. Depending on $D$ and $A$, there exist cases with $X < 1$.

To prove this claim, we need to show that when $M$ is \acs{CAM} or \acs{MM}, $\exists \ D, A$ such that:
\begin{equation}
\label{eq:max}
\max_{r \in D^*}M(\hat{r}_t) < 1
\end{equation}
i.e., for each ranking of documents in $D^*$ the maximum measure score will be less than $1$. To build such an example, the set $D$ needs to contain documents with not comparable tuples of labels: 
\begin{multline}
\label{eq:max_value_iff}
\exists d_{1}, d_{2} \in D \colon \textrm{GT}(d_{1}) \not\sqsubseteq \textrm{GT}(d_{2}) \textrm{ and } \textrm{GT}(d_{2}) \not\sqsubseteq \textrm{GT}(d_{1}) \iff \\
\exists d_{1}, d_{2} \in D \textrm{ and }\exists a_{1}, a_{2} \in A \colon \\ \textrm{GT}_{a_{1}}(d_{1}) \prec_{a_{1}} \textrm{GT}_{a_{1}}(d_{2}) \textrm{ and } \textrm{GT}_{a_{2}}(d_{2}) \prec_{a_{2}} \textrm{GT}_{a_{2}}(d_{1})
\end{multline}
In this case, \acs{CAM} and \acs{MM} cannot achieve a score equal to $1$ 
as illustrated by the following example.
%
%

Consider the example in $\S$\ref{subsec:example} with $A = \{$relevance, correctness$\}$. 
Let $D$ be a set with $3$ documents $D = \{d_{1}, d_{2}, d_{3}\}$ with labels $\textrm{GT}(d_{1}) = (\texttt{mr}, \texttt{c})$, $\textrm{GT}(d_{2}) = (\texttt{hr}, \texttt{pc})$ and $\textrm{GT}(d_{3}) = (\texttt{hr}, \texttt{nc})$.
Documents $(d_1, d_2)$ and $(d_1, d_3)$ are not comparable and there is no unequivocal way of sorting them, e.g.,
it is not clear if $d_1$ should be ranked before $d_2$ or vice-versa.

Let us consider \ac{CAM} and \ac{MM} instantiated with \ac{AP} and \ac{nDCG}. For \ac{AP} we use a harsh mapping for relevance and correctness, i.e., $\{\texttt{fr}, \texttt{hr}\} \mapsto 1$ and $\{\texttt{mr}, \texttt{nr}\} \mapsto 0$, and $\texttt{c} \mapsto 1$ and $\{\texttt{pc}, \texttt{nc}\} \mapsto 0$. 
%
For \ac{nDCG} we map each category to a different integer, for relevance we have: $\texttt{hr} \mapsto 15$, $\texttt{fr} \mapsto 10$, $\texttt{mr} \mapsto 5$, $\texttt{nr} \mapsto 0$, and for correctness we have: $\texttt{c} \mapsto 10$, $\texttt{pc} \mapsto 5$, $\texttt{nc} \mapsto 0$. The \ac{nDCG} ideal ranking~\citep{JarvelinKekalainen2002} 
for relevance is: $(\texttt{hr}, \texttt{hr}, \texttt{mr})$ and for correctness is $(\texttt{c}, \texttt{pc}, \texttt{nc})$. The \ac{nDCG} $\log$ base is set to $2$. 

For \ac{TOMA} we use the embedding of the first row in Tab.~\ref{tab:example} and as weight function we map each equivalence class to a different integer with step $1$. We instantiate \ac{TOMA} with \ac{AP} and \ac{nDCG} with log base $2$ (Tab.~\ref{tab:AP_MM_CAM}). Since \ac{AP} does not handle multi-graded weights, we map the top half of the equivalence classes to $1$ and the rest to $0$. Tab.~\ref{tab:AP_MM_CAM} shows \ac{CAM}, \ac{MM}, and \ac{TOMA} scores instantiated with \ac{AP} \ac{nDCG} for each possible ranking of documents in $D$.

In Tab.~\ref{tab:AP_MM_CAM} none of the rankings in $D^*$ can achieve a score equal to $1$ for \ac{CAM} and \ac{MM}, while \ac{TOMA} has at least one ranking with score $1$. In \ac{CAM} and \ac{MM} this happens because, any way we sort the documents, either we penalize correctness, e.g., $(d_2, d_3, d_1)$ or we penalize relevance, e.g., $(d_1, d_2, d_3)$. \ac{TOMA} does not have this problem, since it first defines how to sort tuples of labels, then weights them accordingly and computes the measure score. Thus, if we sort documents in the order induced by $\preceq_{*}$, we obtain a score equal to $1$ (proof in the appendix). Experiments on real data confirm this, as detailed next. 

\paragraph{Estimating \ac{CAM} and \ac{MM} Upper Bound}
With the following experiment we show that with real data \ac{CAM} and \ac{MM} can be upper bounded by a value $X$ lower than $1$. 
To estimate the value $X$ with real data, we generate different ideal rankings of documents with different strategies. 
The intuition is that by ranking documents in the best possible way, we should achieve a score equal to $1$, as it happens for any single-aspect evaluation measure computed against the ideal ranking. 
Since \ac{CAM} and \ac{MM} do not define how to sort documents, i.e., a total order relation $\preceq_{*}$, we need to test different possible strategies to build these ideal rankings.

First, we define the ideal rankings obtained with a recursive strategy: these are the ideal rankings for each aspect when considered separately, 
e.g., for $3$ aspects, $a_1$, $a_2$ and $a_3$, with a preference order where $a_1$ is followed by $a_2$, followed by $a_3$: (1) we sort the documents with decreasing label for $a_1$; (2) among the documents with the same label for $a_1$, we sort the documents with decreasing label for $a_2$; (3) among the documents with the same label for $a_1$ and $a_2$, we sort the documents with decreasing label for $a_3$. 
We generate these ideal rankings for each possible preference order among the aspects. 

We also generate $3$ additional ideal rankings: (1) we sum the weights across aspects and sort the documents by this sum; (2) we sum the squared weights across aspects and sort the documents by this sum; (3) we consider the highest weight across aspects and sort documents by their highest weight regardless of the aspect.



Fig.~\ref{fig:plot_measures_ranges_cam} reports the distributions of \ac{CAM} with \ac{AP} scores for the ideal rankings for the Decision Track 2019. These distributions depend on the aspect and the topic. 
We see that the upper bound $X$ is variable and depends on the topic: just for $2\%$ of topics it is equal to $1$ and for $26\%$ topics it is lower than $0.9$.
We obtain similar or even more extreme distributions of scores for all the other tracks (except for Misinformation 2020, reported in the online appendix).


\paragraph{Interpretability of \ac{CAM} and \ac{MM} scores}

Problem $2$ is especially important because it affects the interpretability of \ac{CAM} and \ac{MM} scores.
When a measure is used to assess the quality of a single ranking in isolation, it should be intuitively interpretable~\citep{KumarV10}, 
e.g., \ac{nDCG}=$0.6$ has the intuitive interpretation that the ranking can be further improved by $0.4$. 
If \ac{TOMA} is instantiated with \ac{nDCG}, the intuitive interpretability of \ac{nDCG} holds, but if \ac{CAM} or \ac{MM} are instantiated with \ac{nDCG}, the intuitive interpretability of \ac{nDCG} is lost: by the arguments above, \ac{CAM} and \ac{MM} may fail to obtain an optimal score of $1$, and the optimal score depends on $A$ and $D$, hence it cannot in general be known a priori.

This issue is important for \ac{MM}, which is affected also by Problem $1$, and therefore may have $X << 1$. 
Thus \ac{MM} scores can be compressed towards $0$, and this can lead to cases with many ties, where it is hard to distinguish between different rankings.

\subsection{Experimental Findings}
\label{subsec:findings}

\begin{table*}[tb]
\caption{Kendall's $\tau$ correlation between rankings of systems and discriminative power (the higher, the better; best is in bold). Not all aspect combinations occur in all tracks (marked grey).}
\label{tab:corr}
\resizebox{\textwidth}{!}{%
\begin{tabular}{l|rr|rr|rr|rr|rr|rr|rr|rr|rr|rr}
\toprule
           &\multicolumn{2}{c|}{WEB2009}&\multicolumn{2}{c|}{WEB2010}&\multicolumn{2}{c|}{WEB2011}&\multicolumn{2}{c|}{WEB2012}&\multicolumn{2}{c|}{WEB2013}&\multicolumn{2}{c|}{WEB2014}&\multicolumn{2}{c|}{TASK15}&\multicolumn{2}{c|}{TASK16}&\multicolumn{2}{c|}{DECISION19}&\multicolumn{2}{c}{MISINFO 2020} \\
&NDCG&AP&NDCG&AP&NDCG&AP&NDCG&AP&NDCG&AP&NDCG&AP&NDCG&AP&NDCG&AP&NDCG&AP&NDCG&AP\\
\midrule
&\multicolumn{20}{c}{CORRELATION}\\
\midrule
EUCL - CAM&0.25&0.16&0.18&0.12&0.21&0.11&0.31&0.26&0.22&0.12&0.30&0.23&0.68&0.54&0.63&0.55&0.76&0.60& 0.72 & 0.51 \\
EUCL - MM&0.07&0.04&0.05&0.00&0.06&0.03&0.16&0.13&0.12&0.04&0.17&0.04&0.36&0.17&0.02&-0.10&0.46&0.31 & 0.45 & 0.27 \\
MANH - CAM&0.22&0.16&0.21&0.12&0.16&0.11&0.34&0.26&0.31&0.12&0.41&0.23&0.62&0.54&0.60&0.55&0.69&0.60 & 0.72&0.51\\
MANH - MM&0.06&0.04&0.01&0.01&0.06&0.03&0.16&0.13&0.11&0.04&0.15&0.04&0.28&0.17&-0.06&-0.10&0.41 &0.31&0.45 &0.27\\
CHEB - CAM&0.01&0.02&0.06&0.05&0.02&0.02&0.19&0.15&0.11&0.00&0.19&0.07&0.26&0.16&-0.13&-0.18&0.27 &0.27&0.29&0.24\\
CHEB - MM&0.06&0.09&0.00&0.03&0.09&0.01&0.19&0.16&0.12&0.08&0.14&0.05&0.88&0.86&0.52&0.53&0.58&0.60&0.54&0.52 \\
\midrule
EUCL - MANH&0.36&1.00&0.19&1.00&0.21&1.00&0.34&1.00&0.20&1.00&0.34&1.00&0.87&1.00&0.71&1.00&0.72 &1.00&1.00 &1.00\\
EUCL - CHEB&0.01&0.02&0.10&0.03&0.01&0.03&0.32&0.11&0.22&0.04&0.30&0.12&0.33&0.19&-0.21&-0.21&0.28&0.21 &0.26 &0.20\\
MANH - CHEB&0.01&0.02&0.03&0.03&0.02&0.03&0.21&0.11&0.10&0.04&0.18&0.12&0.32&0.19&-0.24&-0.21&0.25 &0.21 &0.26 &0.20\\
CAM - MM&0.10&0.05&0.05&0.01&0.11&-0.01&0.23&0.13&0.16&0.03&0.26&0.04&0.30&0.18&0.09&0.00&0.51&0.41 &0.51&0.42 \\
\midrule
&\multicolumn{20}{c}{CORRELATION}\\
\midrule
Relevance - Popularity  &0.03&0.05&0.01&0.0&0.01&0.02&0.09&0.09&0.06&0.01&0.07&0.02&0.04&0.04&-0.03&0.01&\cellcolor{gray}&\cellcolor{gray}&\cellcolor{gray}&\cellcolor{gray}\\
Relevance - Non-spam  &0.05&0.03&0.02&0.0&0.03&0.01&0.07&0.05&-0.02&-0.01&0.07&-0.01&0.25&0.17&-0.07&-0.08&\cellcolor{gray}&\cellcolor{gray}&\cellcolor{gray}&\cellcolor{gray}   \\
Popularity - Non-spam &0.04&0.03&0.02&-0.01&0.04&0.01&0.07&0.04&-0.03&-0.02&0.04&0.00&0.07&0.02&0.08&0.06& \cellcolor{gray} &\cellcolor{gray}&\cellcolor{gray}&\cellcolor{gray} \\
Relevance - Usefulness  &\cellcolor{gray}&\cellcolor{gray}&\cellcolor{gray}&\cellcolor{gray}&\cellcolor{gray}&\cellcolor{gray}&\cellcolor{gray}&\cellcolor{gray}&\cellcolor{gray}&\cellcolor{gray}&\cellcolor{gray}&\cellcolor{gray}&0.75&0.75&0.75&0.74&\cellcolor{gray}&\cellcolor{gray} &\cellcolor{gray}&\cellcolor{gray}  \\
Usefulness - Popularity &\cellcolor{gray}&\cellcolor{gray}&\cellcolor{gray}&\cellcolor{gray}&\cellcolor{gray}&\cellcolor{gray}&\cellcolor{gray}&\cellcolor{gray}&\cellcolor{gray}&\cellcolor{gray}&\cellcolor{gray}&\cellcolor{gray}&0.10&0.06&-0.04&0.00&\cellcolor{gray}&\cellcolor{gray} &\cellcolor{gray}&\cellcolor{gray}  \\
Usefulness - Non-spam &\cellcolor{gray}&\cellcolor{gray}&\cellcolor{gray}&\cellcolor{gray}&\cellcolor{gray}&\cellcolor{gray}&\cellcolor{gray}&\cellcolor{gray}&\cellcolor{gray}&\cellcolor{gray}&\cellcolor{gray}&\cellcolor{gray}&0.40&0.33& -0.16&-0.19&\cellcolor{gray}&\cellcolor{gray} &\cellcolor{gray}&\cellcolor{gray}  \\
Credibility - Correctness &\cellcolor{gray}&\cellcolor{gray}&\cellcolor{gray}&\cellcolor{gray}&\cellcolor{gray}&\cellcolor{gray}&\cellcolor{gray}&\cellcolor{gray}&\cellcolor{gray}&\cellcolor{gray}&\cellcolor{gray}&\cellcolor{gray}&\cellcolor{gray}&\cellcolor{gray}&\cellcolor{gray}&\cellcolor{gray}&0.26&0.26&0.28&0.24 \\
Relevance - Credibility &\cellcolor{gray}&\cellcolor{gray}&\cellcolor{gray}&\cellcolor{gray}&\cellcolor{gray}&\cellcolor{gray}&\cellcolor{gray}&\cellcolor{gray}&\cellcolor{gray}&\cellcolor{gray}&\cellcolor{gray}&\cellcolor{gray}&\cellcolor{gray}&\cellcolor{gray}&\cellcolor{gray}&\cellcolor{gray}&0.33&0.33 &0.29 &0.25  \\
Relevance - Correctness &\cellcolor{gray}&\cellcolor{gray}&\cellcolor{gray}&\cellcolor{gray}&\cellcolor{gray}&\cellcolor{gray}&\cellcolor{gray}&\cellcolor{gray}&\cellcolor{gray}&\cellcolor{gray}&\cellcolor{gray}&\cellcolor{gray}&\cellcolor{gray}&\cellcolor{gray}&\cellcolor{gray}&\cellcolor{gray}&0.42&0.49  &0.49 &0.47 \\
\midrule
&\multicolumn{20}{c}{DISCRIMINATIVE POWER OF MEASURES}\\
\midrule
CAM &75.98&64.43&66.32&61.23&75.14&61.64&\bf{68.71}&56.74&76.89&57.05&85.06&78.85&53.33&33.33&72.22&55.56&72.58&70.56 &71.53&70.90  \\
MM &75.61&50.58&\bf{72.89}&\bf{67.79}&67.32&67.81&62.68&56.12&\bf{80.71}&46.99&74.25&53.56&0.00&0.00&0.00&0.00&60.08&53.23&68.31&62.20\\
EUCL &75.29&72.64&62.96&66.75&75.14&70.33&66.13&64.10&75.14&\bf{59.45}&80.92&78.85&\bf{66.67}&\bf{66.67}&69.44&\bf{75.00}&73.59&\bf{73.99}&72.86&\bf{75.14}\\
MANH &\bf{76.66}&\bf{72.68}&63.59&67.14&\bf{77.32}&\textbf{70.38}&66.05&\bf{64.18}&76.67&59.34&\bf{86.44}&\bf{79.08}&\bf{66.67}&53.33&\bf{75.00}&\bf{75.00}&\bf{73.79} &73.79&\textbf{73.02}&74.98 \\
CHEB &50.18&6.32&59.82&51.49&73.06&50.11&61.08&39.36&77.10&49.34&75.17&66.21&0.00&0.00&0.00&0.00&42.54 &29.84&65.41&53.33 \\
\bottomrule
\end{tabular}
}
\end{table*}

Empirically, evaluation measures are commonly assessed in terms of their correlation~\cite{Ferrante2019}, discriminative power~\cite{sakaibootstrap}, informativeness~\cite{AslamYP05}, 
intuitiveness \cite{Sakai2012}, and unanimity~\cite{AlbahemEtAl2019}. Out of these, we report only correlation and discriminative power because the rest does not apply: the informativeness test~\cite{AslamYP05} 
requires a precision recall-curve, which cannot be defined for multi-aspect evaluation; 
the intuitiveness test \cite{Sakai2012} requires simple single-aspect measures (e.g.~precision, recall), which do not apply to multi-aspect evaluation; 
the unanimity test~\cite{AlbahemEtAl2019}, which is defined for multi-aspect evaluation, requires that all the simple measures agree over all aspects, which happened extremely rarely in our data, especially as the number of aspects increased (see the low correlation among aspects in Tab.~\ref{tab:corr}). 



\subsubsection{Correlation Analysis}
\label{ss:correlationanalysis}
We use Kendall's $\tau$~\cite{Kendall1945} to estimate \ac{TOMA}'s correlation to \ac{MM} and \ac{CAM}. Generally, if a new evaluation measure \textit{strongly} correlates to an existing one, it is likely to represent redundant information~\cite{Webber:2008:PCR:1390334.1390456}. We use Kendall's $\tau$ because it has better gross-error sensitivity than the Pearson correlation coefficient~\cite{CrouxAndDehon2010}, and because the Spearman correlation coefficient cannot handle ties. 
As per~\cite{Ferrante2019}, we compute the correlation topic-by-topic. For each topic we consider the \ac{RoS} corresponding to two different measures (one ranking per measure) and then compute Kendall's $\tau$ between the two \ac{RoS}. We report Kendall's $\tau$ averaged across all topics. 
As per~\cite{Voorhees1998,Voorhees2001}, we consider two rankings equivalent if Kendall's $\tau$ is greater than $0.9$. 

Tab. \ref{tab:corr} shows the findings, which are summarised as follows:
\begin{itemize}
    \item The \ac{RoS} corresponding to EUCL - MANH are equivalent ($\tau=1$) at all times for \ac{AP}. 
    This perfect correlation for \ac{AP} happens because, by definition, when the sets of equivalence classes from these approaches are mapped to binary labels, they produce the exact same set of labels (see also Tab.~\ref{tab:AP_MM_CAM}). For \ac{nDCG}, $\tau=0.19-1$, where higher correlations correspond to tracks where some aspects are not assessed for non relevant documents, thus there are less extreme cases and EUCL is more similar to MANH.
    \item The \ac{RoS} corresponding to (EUCL, MANH) - CHEB are very weakly correlated ($\tau = 0.01 -  0.32$), this is due to Chebyshev distance being very harsh, since many equivalence classes are considered equivalent to the class of non relevant documents.
    \item The \ac{RoS} corresponding to EUCL - CAM and MANH - CAM are very weakly correlated ($\tau=0.11-0.41$) for the Web tracks, but moderately correlated ($\tau=0.54-0.76$) for the Task, Decision and Misinformation tracks. This happens because: (i) the runs submitted to the Web tracks were not designed to account for multiple aspects and (ii) for the Task, Decision and Misinformation tracks, usefulness, credibility and correctness are not assessed for non relevant documents. Therefore, since some of the values are missing, these methods generate a lower number of equivalence classes, which make them more similar to \ac{CAM}. Whereas, for the Web tracks, popularity and non-spamminess are approximated for all documents, meaning that MANH and EUCL can possibly generate all the different equivalence classes, even for non relevant documents. This makes them less similar to \ac{CAM} than on the Task or Decision tracks.
    \item For the Task, Decision and Misinformation tracks, the \ac{RoS} corresponding to \ac{MM} and CHEB are moderately correlated ($\tau=0.52-0.88$). The fact that, for these tracks, usefulness, credibility and correctness are not assessed for non relevant documents, means that all the documents that are mapped to a $0$ weight with CHEB, are also contributing as $0$ to \ac{MM}.
\end{itemize}

To contextualise these findings, the middle part of Tab. \ref{tab:corr} shows the $\tau$ values of the \ac{RoS} corresponding to evaluating a single aspect only. Overall, the resulting correlations are low to non-existent, meaning that considering multiple aspects affects the final evaluation outcome. The two exceptions where the correlation between \ac{RoS} is not very low are:
\begin{itemize}
    \item For Task 2015-2016, for relevance - usefulness, $\tau=0.74-0.75$. This happens because: (1) usefulness is not assessed for non relevant documents, thus non relevant documents are assumed to be not useful, and (2) usefulness is a very sparse signal (1.75\% of documents are useful).
    \item For the Decision and Misinformation Tracks, for all aspects, $\tau=0.24-0.49$. Again here credibility and correctness are not assessed for non relevant documents ($6.89\%$ of documents are credible and $9.75\%$ are correct for the Decision Track; $13.73\%$ of documents are correct and $27.62\%$ are credible for the Misinformation Track), so the correlation is not as high as for the Task tracks.
\end{itemize}


Overall, the most correlated \ac{RoS} correspond to: EUCL - MANH ($\tau$ up to $1$), (EUCL, MANH)- CAM ($\tau$ up to $0.76$), and CHEB - MM ($\tau$ up to $0.88$).
Intuitively, EUCL and MANH may be more similar to \ac{CAM} (mean), while CHEB may be more similar to \ac{MM} (harmonic mean). 
Thus \ac{TOMA} proposes an alternative evaluation framework, which overcomes \ac{CAM} and \ac{MM} anomalies (see $\S$\ref{subsec:anomalies}).
The fact that $\tau$ values between \ac{TOMA} and the baselines are never above $0.9$ means that there are noticeable differences between the \ac{RoS} generated by \ac{TOMA} and by \ac{CAM} or \ac{MM}. 
Recall that all measures are instantiated with NDCG or AP, meaning that differences between them are due to how multi-aspect labels are treated.

\subsubsection{Discriminative Power}
\label{subsec:discr}

We use Bootstrap Hypothesis Test~\cite{sakaibootstrap} to estimate the discriminative power of TOMA, CAM and MM. Given a set of topics and a set of runs, we first generate subsets of topics by sampling with replacement the complete set of topics. We set the number of bootstrap samples to $10\,000$. To assess whether the measure scores for pairs of runs can be considered different at a given confidence level, we use a Paired Bootstrap Hypothesis Test. The confidence level is $1-\alpha$, where $\alpha$ is the Type I Error, i.e., the probability to consider two systems different even if they are equivalent. We set 
$\alpha=0.01$, requiring 
strong evidence for two systems to be different.

Tab.~\ref{tab:corr} (bottom part) displays the results of the discriminative power analysis, where 
the higher the score, the more discriminative  (i.e., the better) the approach. We see that $16/20$ times either MANH ($12/20$) or EUCL ($6/20$)\footnote{Ties are included in these counts.} is best. The remaining $4$ times, \ac{MM} is best $3$ times, and \ac{CAM} once. We also see that CHEB is never best, and for Task 2015-2016 it is actually zero. This is due to the very small amount of positive labels for usefulness in that track. For the same reason, \ac{MM} is also zero for the same track. Overall, CHEB is the least discriminative measure, followed by \ac{MM}; this is due to how these methods treat tuples of labels: the fact that if one aspect label is zero, then the whole score is zero, practically means that many runs are considered equal purely on that basis.

\subsubsection{Zero-aspect documents}
\label{subsec:zero}


\begin{table}[tb]
\centering
\caption{Number of times (\%) that the labels of all aspects sum to $0$ for a document that is ranked at position $1$-$5$ (column 1) in a run that has been assessed as best per $\{$topic, track, year$\}$ separately with $\{$\ac{CAM}, \ac{MM}, EUCL, MANH, CHEB$\}$ using a retrieval cutoff of 5. The lower, the better.}
\label{tab:zero}
\resizebox{.47\textwidth}{!}{%
\begin{tabular}{c|rrrrr}
\toprule
Rank &CAM&MM&EUCL&MANH&CHEB \\ 
\midrule
1 &51 (1.18\%) &131 (3.02\%) &39 (0.90\%) &\bf{33 (0.76\%)} &154 (3.55\%)\\
2 &65 (1.50\%) &159 (3.67\%) &50 (1.15\%) &\bf{48 (1.11\%)} &179 (4.13\%)\\
3 &103 (2.38\%) &202 (4.66\%) &88 (2.03\%) &\bf{78 (1.80\%)} &185 (4.17\%)\\
4 &102 (2.35\%) &173 (3.99\%) &86 (1.99\%) &\bf{74 (1.71\%)} &183 (4.23\%)\\
5 &107 (2.47\%) &196 (4.53\%) &95 (2.19\%) &\bf{81 (1.87\%)} &205 (4.73\%)\\
1-5 &428 (9.88\%) &861 (19.88\%) &358 (8.27\%) &\bf{314 (7.25\%)} &906 (20.92\%)\\
\bottomrule
\end{tabular}%
}
\end{table}
Our next analysis is motivated by the empirical trash$@k$ measure often used in industry to mitigate the high cost of retrieving ``trash'' in high ranks. We count how often the labels of all aspects sum to zero for a document that has been ranked at position $1$-$5$ in a run that has been assessed as the best run per track year, on a per query basis, using a retrieval cutoff of $5$, separately with $\{$\ac{CAM}, \ac{MM}, EUCL, MANH, CHEB$\}$  when instantiated separately with \ac{nDCG} and \ac{AP}. When the labels of all aspects sum to zero, this means that the corresponding document is of the worst quality. Ideally, such documents should not be retrieved, but when they do, they should not be in the top 5. 

In Tab. \ref{tab:zero} we see that MANH is associated with the lowest amount of zero-aspect documents, closely followed by EUCL. This happens because MANH is designed so that the higher the sum of a document's labels across aspects, the better that document will be deemed. CHEB is overall worst, closely followed by \ac{MM}. This closeness between EUCL-MANH and CHEB-\ac{MM} agrees with the previous correlation and discriminative power analysis. Overall, MANH (and less so EUCL) penalise low quality documents the best.

\subsubsection{Document quality @1-100}
\label{subsec:}

\begin{table}[tb]
\centering
\caption{Average sum of aspect labels for a document that is ranked at position $1$-$100$ (column 1) in a run that has been assessed as best per $\{$topic, track, year$\}$ separately with $\{$\ac{CAM}, \ac{MM}, EUCL, MANH, CHEB$\}$ using a retrieval cutoff of 100. The higher, the better.}
\label{tab:quart}
\begin{tabular}{c|ccccc}
\toprule
Ranks &CAM&MM&EUCL&MANH&CHEB \\ 
\midrule
1-25&\bf{1.70}&1.49&1.67&1.69&1.39\\
26-50&0.85&0.78&0.91&\bf{0.94}&0.70\\
51-75&0.57&0.53&0.63&\bf{0.64}&0.48\\
76-100&0.40&0.39&0.43&\bf{0.44}&0.36\\
\bottomrule
\end{tabular}%
\end{table}

We look at the quality of documents that have been ranked at positions 1-100 in a run that has been assessed as best per $\{$topic, track, year$\}$ separately with $\{$\ac{CAM}, \ac{MM}, EUCL, MANH, CHEB$\}$, when instantiated separately with \ac{nDCG} and \ac{AP}, using a retrieval cutoff of $100$. We split the ranks $1$-$100$ into four sets ($1$-$25$, $26$-$50$, $51$-$75$, $76$-$100$). For each document in each set, we sum the labels of its aspects, and we report the average of these sums, which can be seen as an approximation of the average document quality (the higher, the better). 

As expected, we see that the numbers in Tab. \ref{tab:quart}, and hence document quality, drop as we move down the ranking, at all times. Comparing across columns however, we see that, for the runs that were assessed as best by MANH, document quality is overall, albeit marginally, the best, at ranks 26-100. This illustrates that the design of MANH (the higher the sum of a document's labels across aspects, the better that document will be considered) gives it the practical advantage of, not only reducing the amount of low quality documents in the top ranks (as seen in Tab. \ref{tab:zero}), but also of increasing the quality of documents further down the ranking, as we see now. Again, as previously, we observe that EUCL is a close second-best method, CHEB and MM are overall worst, and CAM is in between (although best, together with MANH, for the top ranks).

\section{Conclusion and Limitations}
\label{sec:conclusion}
\textit{Multi-aspect evaluation} is a special case of \ac{IR} evaluation where the ranked list of documents returned by an \ac{IR} system in response to a query must be assessed in terms of not only relevance, but also other \textit{aspects} (or dimensions) of the ranked documents, e.g., credibility or usefulness. We presented a principled multi-aspect evaluation approach, called \ac{TOMA}, that is defined for any number and type of aspect, and that allows for (i) aspects having different gradings, (ii) any relative importance weighting for different aspects, and (iii) integration with any existing single-aspect evaluation measure, such as \ac{nDCG}. We showed that \ac{TOMA} has better discriminative power than prior approaches to multi-aspect evaluation, and that it is better at rewarding high quality documents across the ranking. 

One limitation of \ac{TOMA} is represented by the arbitrary choices of the embedding function, the distance function and the weight function.
The embedding function maps labels from a nominal or ordinal scale to an interval or ratio scale.
This calls for a in-depth investigation of the theoretical properties of \ac{TOMA} using the existing axiomatic treatments
of \ac{IR} effectiveness measures~\cite{AmigoEtAl2018c,MaddalenaMizzaro2014,BusinMizzaro2013,AmigoAndMizzaro2020,FerranteEtAl2021}.
This also motivates a deep analysis of the interactions between different aspects and/or documents and how to handle them with \ac{TOMA}, for example by defining a proper embedding and distance function which account for aspects as diversity, novelty, and redundancy.
Moreover, the embedding function combined with the distance function can generate a large number of tuples of labels, which can be mapped to different integers through the weight function.
This might be a problem for gain based measures, thus a possible solution is to use \ac{TOMA} to define the ideal ranking and then use effectiveness measures based on similarity to ideal rankings ~\cite{DBLP:conf/ictir/ClarkeVS20,DBLP:conf/cikm/ClarkeSV20,ClarkeEtAl2021}.
Finally, the empirical impact of varying both distance and weight functions should also be investigated, as should
the impact of employing further multi-graded measures as $\acs{ERR}$~\cite{ChapelleEtAl2009}, and the alignment of our current approach with real user preferences.

\vspace{0.5em}
\small
\noindent \textbf{Acknowledgments}. This paper is partially supported by the EU Horizon 2020 research and innovation programme under the MSCA grant No. 893667.

\clearpage
\bibliographystyle{ACM-Reference-Format}
\bibliography{bibliography,zz-proceedings}

\end{document}